\documentclass[twocolumn,showpacs,showkeys,preprintnumbers,amsmath,amssymb,pra,floatfix]{revtex4}
\usepackage{graphicx}
\usepackage{dcolumn}
\usepackage{bm}
\usepackage{psfrag}

\begin{document}
\title{Non-Hermitian Quantum  Annealing in the Ferromagnetic Ising Model }
\author{Alexander I. Nesterov}
   \email{nesterov@cencar.udg.mx}
\affiliation{Departamento de F{\'\i}sica, CUCEI, Universidad de Guadalajara,
Av. Revoluci\'on 1500, Guadalajara, CP 44420, Jalisco, M\'exico}
\author{Gennady P.  Berman}
 \email{gpb@lanl.gov}
\affiliation{Theoretical Division, Los Alamos National Laboratory,
Los Alamos, NM 87544, USA}
\author{Juan Carlos Beas Zepeda}
   \email{juancarlosbeas@gmail.com}
\affiliation{Departamento de F{\'\i}sica, CUCEI, Universidad de Guadalajara,
Av. Revoluci\'on 1500, Guadalajara, CP 44420, Jalisco, M\'exico}
\date{\today}
\begin{abstract}
We developed a non-Hermitian quantum optimization algorithm to find the ground state of the ferromagnetic Ising model with up to 1024 spins (qubits). Our approach leads to significant reduction of the annealing time. Analytical and numerical results  demonstrate that the total annealing time is proportional to $\ln N$, where $N$ is the number of spins. This encouraging result is important in using classical computers in combination with quantum algorithms for the fast solutions of NP-complete problems. Additional research is proposed for extending our dissipative algorithm to more complicated problems.
\end{abstract}


\maketitle

\section{Introduction}

 Quantum annealing (QA) algorithms can be useful for solving many hard problems related to finding the global minimum of multi-valued functions, cost functions, optimal configurations of complex networks, and the ground states of the corresponding Hamiltonians. The main idea of QA is to utilize the collective quantum tunneling effects enabling complex systems to tunnel during the slow time-evolution from local minima to the global ground state. Although many useful results have been obtained in this field, many problems still need to be resolved. The main challenge is to accelerate the speed of QA algorithms so that the annealing time grows not exponentially but polynomially  with the size of the problem \cite{KN,FGGLL,SSMO,DC,SMTC,OMNH,TSTR,SST,SNS,AM}.

There are many approaches to finding the ground state of the Hamiltonian, $\mathcal H_0$, using QA algorithms \cite{KN,FGGLL,SSMO,DC,SMTC,OMNH,TSTR}. One of these approaches is based on introducing a time-dependent Hamiltonian,
$\mathcal H(t) =\mathcal H_0 +  \Gamma(t)\mathcal H_1$, where $\mathcal H_0$ is the Hamiltonian to be optimized, $\mathcal H_1$ is an auxiliary (``initial") Hamiltonian and $[\mathcal H_0,\mathcal H_1] \neq 0$. The term $\Gamma(t)\mathcal H_1$ provides the non-trivial quantum dynamics required during annealing. The external time-dependent field, $\Gamma(t)$, is a control parameter that decreases from a large enough value to zero during the evolution. The ground state of $\mathcal H_1$ is the initial state. If $\Gamma(t)$ decreases sufficiently slowly, the adiabatic theorem guarantees finding the ground state of the main Hamiltonian, $\mathcal H_0$, at the end of computation.

In order to reach the desired ground state, QA algorithms usually require the existence of a finite gap between the ground state and first excited state. However, in typical cases the minimal gap, $g_m$, is exponentially small in the number of qubits, $m$. For instance, in the commonly used quantum optimization $m$-qubit models the minimal energy gap is $g_m \approx 2^{-m/2}$ \cite{FGGLL,DC,SUD,JKKM,YKS}. This causes the annealing time to increase exponentially with the size of the system. Then, the problem arises on how to speed up the performance of the QA algorithms and, at the same time, find the ground state of $\mathcal H_0$.

In  \cite{BN} we proposed an alternate adiabatic quantum optimization algorithm based on the non-Hermitian quantum mechanics. Recently, we applied this non-Hermitian quantum annealing (NQA) algorithm to Grover's problem of finding a marked item in an unsorted database \cite{NABG}. We showed that a search time depends on the chosen relaxation parameters, and is proportional to the logarithm of the number of qubits.

In this paper, we apply the NQA to study a dissipative dynamics in the one-dimensional ferromagnetic Ising spin chain. We assume that the dissipation vanishes at the end of evolution. So, after the annealing is finished, the system is governed by the Hermitian Hamiltonian.

We show that a dissipation significantly increases the probability for the system to remain in the ground state. In particular, a comparison with the results of the Hermitian QA reveals that the NQA reaches the ground state of $\mathcal H_0$ with much larger probability, if we use the same annealing scheme. We show that the NQA has a complexity of order, $\ln N$, where $N$ is the number of spins. This is much better than the quantum Hermitian adiabatic algorithm yielding the complexity of order $N^2$.

A dissipative term which we use corresponds to a tunneling of the system to its own continuum, as usually happens when one applies a Feshbach projection method on intrinsic states in nuclear physics and quantum optics. In our case, the intrinsic states are the states of the quantum computer. So, the absolute probability of our quantum computer to survive during the NQA can be small. That is why we use a ratio of two probabilities -- the probability for the system to remain in the ground state to the probability of survival of the quantum computer. This relative probability (we call it ``intrinsic'' probability) is well-defined, and remains finite during the NQA. So, our approach cannot be directly used in the experiments on QA, but rather as a combination of classical computer and NQA to significantly decrease the time of annealing. Also, a dissipative term which we use in this paper is rather artificial in a sense that it has no a direct relation to a real physical dissipative mechanisms. At the same time, we note  that our dissipation term corresponds, in principal, to the tunneling effects in the superconducting phase qubits if tunneling is realized mostly from the lowest energy levels. Our hope is that a reduction of the calculation annealing time can help to boost solution of the NP-complete problems by using combination of classical computers and NQA algorithms.

This paper is organized as follows. In Sec. II, the generic adiabatic quantum optimization (AQO) algorithm based on non-Hermitian QA is discussed. In Sec. III, we introduce a lossy one-dimensional Ising system in a transverse magnetic field governed by a non-Hermitian Hamiltonian. In Sec. IV, we study the quench dynamics of our system using both analytic and numerical methods. We conclude in Sec. V with a discussion of our results. In the Appendices we present some technical details.

\section{ Non-Hermitian Quantum Annealing: preliminaries}

The generic adiabatic quantum optimization problem based on the QA algorithm can be formulated as follows \cite{SSMO}. Let $\mathcal H_0$ be the Hamiltonian whose ground state is to be found, and $\mathcal H_1$ be an auxiliary ``initial''  Hamiltonian. We consider the following time-dependent Hamiltonian:
\begin{equation}\label{QA1}
\mathcal H_\tau(t) = \mathcal H_0 +  g(t)\mathcal H_1,
\end{equation}
where $[\mathcal H_0,\mathcal H_1] \neq 0$. The function, $g(t)$, is monotonic and satisfies the relation, $g(\tau) = 0$. It is assumed that $\mathcal H_1$ is dominant at the initial time $t=0$, and, since $g(\tau) = 0$, the $\mathcal H_\tau(t) \rightarrow \mathcal H_0$ as $t\rightarrow \tau$.

The evolution of the system is determined by the Schr\"odinger equation:
\begin{eqnarray}\label{Sch1}
i\frac{\partial }{\partial t}|\Psi(t)\rangle = { \mathcal H_\tau}(t)|\Psi(t)\rangle.
\end{eqnarray}
The initial conditions are imposed as follows: $|\Psi(0)\rangle= |\Psi_g\rangle$, where $|\Psi(0)\rangle$ is the ground state of $\mathcal H_1$. The adiabatic theorem guarantees that the initial state, $|\Psi_g\rangle$, evolves into the final state of $|\Psi_g(\tau)\rangle$, which is the ground state of the Hamiltonian, $\mathcal H_0$, as long as the instantaneous ground state of $\mathcal H_\tau(t)$ does not become degenerate at any time.

The validity of the adiabatic theorem requires
\begin{equation}\label{QA2}
    \sum_{m\neq n}\bigg|\frac{\langle \psi_m|\partial \mathcal H_\tau/\partial
    t|\psi_n\rangle}{(E_m - E_n)^2}\bigg|\ll 1.
\end{equation}
This restriction is violated near the degeneracies in which the eigenvalues coalesce.
In the common case of double degeneracy with two linearly independent eigenvectors, the energy surfaces form the sheets of a double cone. The apex of the cones is called the ``diabolic point'', and since, for a generic Hermitian Hamiltonian, the co-dimension of the diabolic point is three, it can be characterized by three parameters  \cite{B0,BW}.
For the quantum optimization governed by the Hamiltonian of Eq. (\ref{QA1}), the requirement (\ref{QA2}) can be written as \cite{SSMO,DC}
\begin{equation}\label{QA3}
    \tau \gg \tau_0 =\frac{\max|\langle \psi_e|\partial\mathcal H_\tau/\partial
    s|\psi_g\rangle|}{\min|E_e - E_g|^2},
\end{equation}
where $s=t/\tau$, and $E_e$ is the energy of the first excited state, $|\psi_e\rangle$. Thus, if at the time, $\tau_c < \tau$, the gap, $\Delta E = |E_e - E_g|$, is small enough, the time required to pass from the initial state to the final state becomes very large, and the AQO loses its advantage over thermal annealing.

Recently \cite{BN}, we proposed a generic  non-Hermitian adiabatic quantum optimization. Here we consider a particular case of the NQA. Let $\mathcal H_0$ be a Hamiltonian whose ground state is to be found, and let,
\begin{align}
\mathcal{ \tilde H}_1(t) = g(1- t/\tau)\mathcal H_1 - i\delta(1- t/\tau)\mathcal H_2,
\end{align}
be the non-Hermitian auxiliary ``initial''  Hamiltonian.

Consider the following time-dependent Hamiltonian:
\begin{equation}\label{NQA4}
 {\mathcal {\tilde H}_\tau(t)} = \mathcal H_0 + \mathcal{ \tilde H}_1(t),
\end{equation}
where $[\mathcal H_0,\mathcal H_1(0)] \neq 0$. We impose the initial conditions as follows: $|\Psi(0)\rangle= \Psi_g\rangle$, so that
$\mathcal{\tilde H}_1|\psi_g\rangle= E_g|\psi_g\rangle$ with $E_g$ being the energy of the ground state of the auxiliary non-Hermitian Hamiltonian $\mathcal{\tilde H}_1$. At the end of evolution the total Hamiltonian ${\mathcal {\tilde H}_\tau(\tau)} = \mathcal H_0 $, and the adiabatic theorem provides that the final state be the ground state of $\mathcal H_0$, if the evolution was slow enough.

We denote by $|\psi_n(t)\rangle$ and $\langle\tilde\psi_n(t)|$ the right and left instantaneous eigenvectors of the total Hamiltonian:
$   {\mathcal {\tilde H}_\tau(t)}|\psi_n (t)\rangle = E_n(t)|\psi_n\rangle$, $\langle\tilde\psi_n (t)| {\mathcal {\tilde H}_\tau(t)}=
    \langle\tilde\psi_n(t)|E_n(t)$.
We assume that these eigenvectors form a bi-orthonormal basis,
$\langle\tilde\psi_m|\psi_{n}\rangle = \delta_{mn}$ \cite{MF}.

For the non-Hermitian quantum optimization problem governed by the Hamiltonian (\ref{NQA4}), the validity of adiabatic approximation requires
\begin{equation}\label{NQA3}
  \tau \gg \frac{\max |\langle \tilde \psi_e |\mathcal {\dot {\tilde H}_\tau}(t)|\psi_g\rangle|}{\min |E_e(t)- E_g(t)|^2},
\end{equation}
where the ``dot'' denotes the derivative with respect to the dimensionless time, $s=t/\tau$. This restriction is violated near the ground state degeneracy, where complex energy levels cross. The point of degeneracy is known as the exceptional point, and it is characterized by a coalescence of eigenvalues and their corresponding eigenvectors, as well. Therefore, studying the behavior of the system in the vicinity of the exceptional point requires a special care \cite{SKM,KMS,MKS1}.

In the vicinity of the level crossing point, only the two-dimensional Jordan block, related to the level crossing, makes the most considerable contribution to the quantum evolution. Then, the multi-dimentional problem can be described by the effective two-dimensional Hamiltonian, acting in the subspace spanned by the ground state and the first excited state of the total non-Hermitian Hamiltonian, ${\tilde H}_\tau$ \cite{BN}.

\section{Description of the model}

In this section, we consider the one-dimensional Ising model in a transverse magnetic field with dissipation governed by the following non-Hermitian Hamiltonian:
\begin{eqnarray} \label{EqH1}
H= - \frac{J}{2}\sum^N_{n=1}\big(g \sigma^x_n + \sigma^z_n \sigma^z_{n+1} + i2{\delta}\sigma^{-}_n \sigma^{+}_n\big),
\end{eqnarray}
with periodic boundary condition, $\mbox{\boldmath$\sigma$}_{N+1} =\mbox{\boldmath$\sigma$}_1$. The external magnetic field is associated with the parameter, $g$, the rate of decay is described by the parameter, $\delta$, and $\sigma^{\pm}_n =  (\sigma^{z}_n \pm i \sigma^{y}_n)/2$ are the spin raising and lowering operators.

In principle, this model can be realized in a chain of the superconducting phase qubits with the tunneling to the continuum mostly from the lowest energy levels, and by applying the standard Feshbach projection method to obtain effective non-Hermitian Hamiltonian. To get the Hamiltonian of Eq. (\ref{EqH1}) one needs to make rotation along the axes $y$ by $\pi/2$ radians.

We apply the standard Jordan-Wigner transformation, and the following procedure outlined in \cite{KSH,DJ,CP,NO},
\begin{align}
\sigma^x_n = 1-2 c^\dagger_n c_n, \\
\sigma^y_n = i(c^\dagger_n - c_n  )\prod_{m<n}(1-2 c^\dagger_m c_m), \\
\sigma^z_n = -(c_n + c^\dagger_n )\prod_{m<n}(1-2 c^\dagger_m c_m),
\end{align}
in which  $c_n$ are fermionic operators that satisfy anticommutation relations:
 $\{c^\dagger_m,c_n\} = \delta_{mn}$ and $\{c_m,c_n\}=\{c^\dagger_m,c^\dagger_n\}=0$. Then, we obtain
 $H= P^+ H^+P^+  +   P^- H^- P^-$, where
 \begin{align}
 P^{\pm} = \frac{1}{2}\Big (1 \pm \prod^N_{n=1}(1-2 c^\dagger_n c_n) \Big)
 \end{align}
denote the projectors onto the subspaces with even $``+ "$ and odd $``-"$ numbers of quasiparticles \cite{KSH,LSM}, and
\begin{eqnarray}\label{I}
H^{\pm} = &-\displaystyle\frac{J}{2}\sum^N_{n=1}\big(c^\dagger_n c_{n+1 }+ c_{n+1} c_{n} + \tilde g + i\delta \nonumber \\
&- 2\tilde g c^\dagger_n c_{n } + c^\dagger_{n +1}c_{n }+ c^\dagger_{n} c^\dagger_{n+1} \big),
\end{eqnarray}
in which $\tilde g = g  + i \delta$.

The Hamiltonian, $H^-$, is related to $c_n$'s with the periodic boundary conditions, $c_{N+1} = c_1$, while in $H^+$ the operators, $c_n$, obey the following (`antiperiodic') boundary conditions: $c_{N+1} =- c_1$. Since the parity of quasiparticles is conserved, one can consider only either $H^+$ or $H^-$. Further we consider only quasiparticles with the even parity.

Applying the Fourier transformations with the antiperiodic boundary condition, $c_{N+1}=-c_1$,
\begin{align}
c_n = \frac{e^{-i\pi/4}}{\sqrt{N}}\sum_k c_k e^{in\varphi_k }, \\
\varphi_k = \frac{2\pi k}{N}, \quad k= \pm \frac{1}{2},\pm \frac{3}{2},\dots,\pm \frac{N-1}{2},
\end{align}
we obtain,  $H^{+}= \sum_{k}H_k$, where
\begin{eqnarray}\label{Ia}
 H_k=\frac{J}{2}\big(2(\tilde g - \cos \varphi_k)c^\dagger_k c_k- \tilde g - i\delta \nonumber \\
+ \sin \varphi_k(c^\dagger_k c^\dagger_{-k} + c_{-k} c_{k} )\big).
\end{eqnarray}

The Hamiltonian, $H^{+}$, can be diagonalized by using the generalized Bogoliubov transformation:
\begin{align}\label{Eq6a}
&c_k = \cos\frac{\theta_k}{2} \, b_k +  \sin\frac{\theta_{-k}}{2} \,b^\dagger_{-k},  \\
& c^\dagger_k =  \cos\frac{\theta_k}{2} \,b^\dagger_k +  \sin\frac{\theta_{-k}}{2} \,b_{-k}, \label{Eq6b}\\
&b_k = \cos\frac{\theta_k}{2} \, c_k +  \sin\frac{\theta_k}{2} \,c^\dagger_{-k}, \label{Eq6c}\\
& b^\dagger_k = \cos\frac{\theta_k}{2} \, c^\dagger_k + \sin\frac{\theta_k}{2} \, c_{-k},\label{Eq6d}
\end{align}
where
\begin{eqnarray}\label{Th}
\cos\theta_k = \frac{\tilde g -  \cos \varphi_k}{\sqrt{\tilde g^2 -  2\tilde g\cos \varphi_k +1}}, \\
\sin\theta_k = \frac{ \cos \varphi_k}{\sqrt{\tilde g^2 -  2\tilde g\cos \varphi_k +1}},
\end{eqnarray}
with $\theta_k$ being a complex angle.

There are two eigenstates for each $k$,
\begin{align}\label{E2a}
 &|u_{+}(k)\rangle = \left(\begin{array}{c}
                 \cos\frac{\theta_k}{2} \\
                  \sin\frac{\theta_k}{2}
                  \end{array}\right), \,
\langle \widetilde u_{+}(k)| = \big(\cos\frac{\theta_k}{2} ,
\sin\frac{\theta_k}{2} \big),\\
&|u_{-}(k)\rangle = \left(\begin{array}{c}
               -\sin\frac{\theta_k}{2}\\
                 \cos\frac{\theta_k}{2}
                  \end{array}\right), \,
\langle \widetilde u_{-}(k)| = \big(-\sin\frac{\theta_k}{2}, \cos\frac{\theta_k}{2}
 \big),
 \end{align}
 with the complex energies, $\varepsilon_{\pm}(k)= -\varepsilon_{0} \pm \varepsilon_k$, where $\varepsilon_{0} = J\cos \varphi_k+ iJ\delta $, and $ \varepsilon_k = J\sqrt{\tilde g^2 -  2\tilde g\cos \varphi_k +1}$.
  \label{E2b}
Here we denote by $ |u_{\pm}(k)\rangle$ ($\langle \widetilde u_{\pm}(k)| $) the right (left) eigenvectors.

With help of Eqs. (\ref{Eq6a}) -- (\ref{Eq6d}) we obtain the diagonalized Hamiltonian as a sum of quasiparticles with half-integer quasimomenta,
\begin{align}\label{H2}
H^{+}&=  -\frac{1}{2}\sum_{k}\varepsilon_0 + \sum_{k} \varepsilon_k\Big( b_k^\dagger b_k  - \frac{1}{2}\Big) \nonumber \\
& =-\sum_{k>0}\varepsilon_0 + \sum_{k>0} \varepsilon_k\Big( b_k^\dagger b_k  -  b_{-k}b_{-k}^\dagger\Big). \end{align}
Its spectrum contains only states with even number of quasiparticles.

The ground state of the Hamiltonian (\ref{H2}) is a state, $|\psi_g\rangle$, annihilated by all quasiparticles annihilation operators, $b_k$, so that, $b_k|\psi_g\rangle =0$. One can show that the ground state can be written as a product of qubit-like states:
\begin{eqnarray}
&|\psi_g\rangle =  \bigotimes_{k} \Big(\cos\frac{\theta_k}{2}|0\rangle_k  |0\rangle_{-k}  -  \sin\frac{\theta_k}{2}|1\rangle_k  |1\rangle_{-k} \Big),  \\
&\langle \tilde\psi_g |=  \bigotimes_{k} \Big(\cos\frac{\theta_k}{2}\langle 0|_k \langle 0|_{-k}  -  \sin\frac{\theta_k}{2}\langle 1|_k \langle 1|_{-k} \Big),
\end{eqnarray}
where $|0\rangle_k $ is the vacuum state of the mode $c_k$, and $|1\rangle_k  $ is the excited state: $|1\rangle_k =c^\dagger_k |0\rangle_k$.

Since for each $k$, the ground state lies into the two-dimensional Hilbert space spanned by $|0\rangle_k  |0\rangle_{-k}$ and $|1\rangle_k  |1\rangle_{-k}$, it is sufficient to project $H_k$ on this subspace. For a given value of $k$, both of these states can be represented as a point on the complex two-dimensional sphere, $S^2_c$. In this subspace the Hamiltonian, $H_k$, takes the form
\begin{align}\label{H1a}
 {H}_k = -{\varepsilon_0} {1\hspace{-.125cm}1} + {J} \left(
\begin{array}{cc}
              \tilde g- \cos \varphi_k & \sin \varphi_k \\
             \sin \varphi_k & -\tilde g+ \cos \varphi_k  \\
            \end{array}
          \right).
\end{align}

For $|\tilde g|\gg 1$, the ground state is paramagnetic with all spins oriented along the $x$ axis, and  from Eq. (\ref{Th}) we obtain $\cos\theta_k \rightarrow 1$ as $|\tilde g| \rightarrow \infty$. Thus, the south pole of the complex Bloch sphere corresponds to the paramagnetic ground state. On the other hand, when $|g|\ll 1$ there are two degenerate ferromagnetic ground states with all spins polarized either up or down along the $z$ axis. The real part of the complex energy reaches its minimum at the point defined by $\cos\theta_k = -1$, and, hence, the north pole of the complex sphere is related to the  pure ferromagnetic ground state with the broken symmetry in which all spins have orientation either up or down. However, in the thermodynamic limit the system passing through the critical point ends in a superposition of up and down states with finite domains of spins separated by kinks \cite{DJ}.

The ground state energy is given by,
\begin{eqnarray}\label{B2}
 E_{gs} = -N\varepsilon_0 - \sum_{k >0}\varepsilon_k,
\end{eqnarray}
and in the thermodynamic limit ($N\rightarrow \infty$) the ground state energy per spin can be written as
\begin{eqnarray}\label{B2a}
 \varepsilon_{gs} = -\varepsilon_0 - \frac{J}{\pi}\int_0^\pi \sqrt{\tilde g^2 -  2\tilde g\cos \vartheta +1}\,d\vartheta,
\end{eqnarray}
where $ \varphi_k \rightarrow \vartheta  $ as $N\rightarrow\infty$. This integral can be written in terms of a complete elliptic integral of the second kind,
\begin{eqnarray}\label{B2b}
 \varepsilon_{gs} = -iJ\delta -\frac{2J(\tilde g+1)}{\pi}\mathbf E \big(2\sqrt{\tilde g}/(\tilde g+1)\big).
\end{eqnarray}

\begin{figure}[tbp]
\scalebox{0.15}{\includegraphics{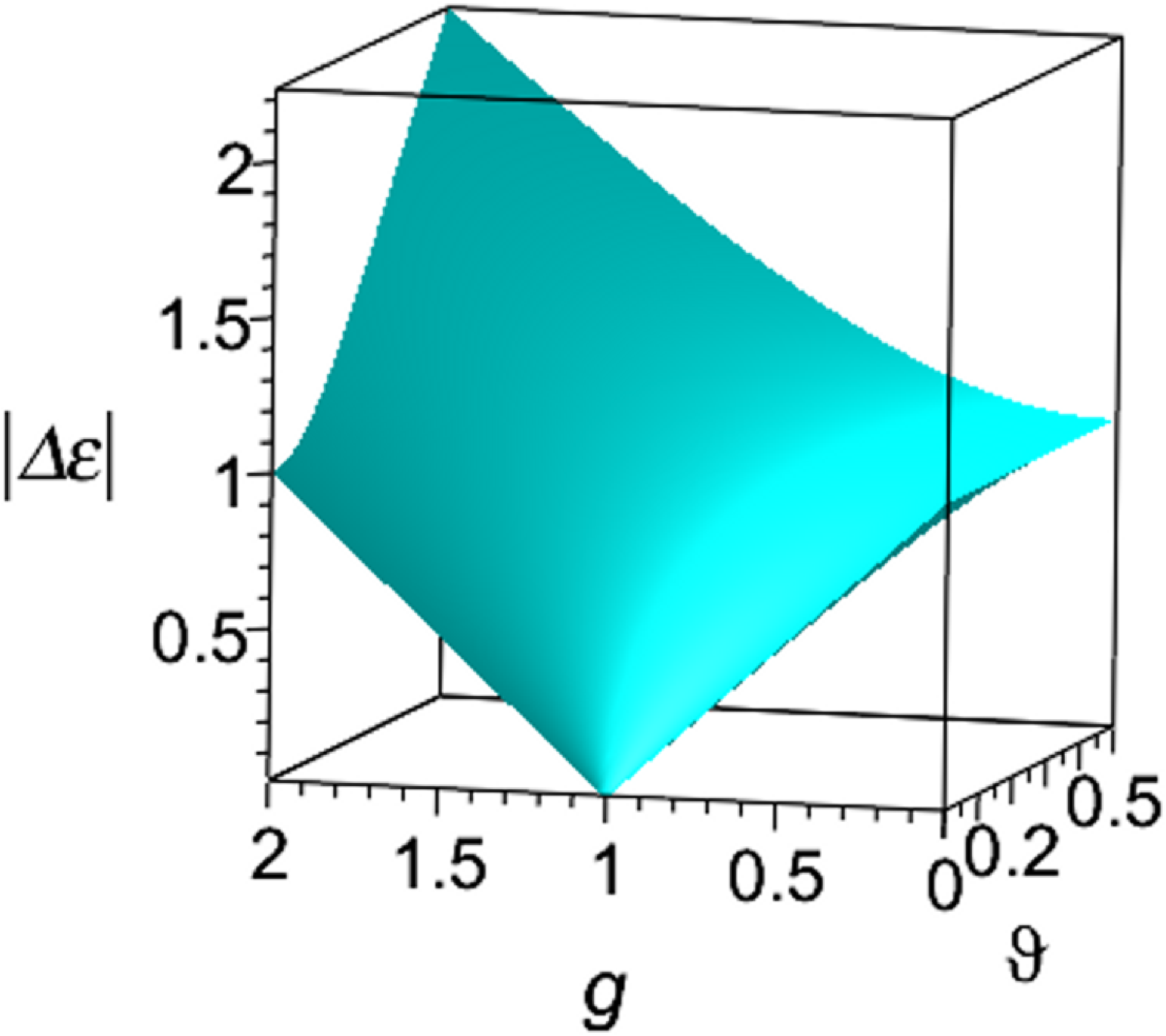}}
\scalebox{0.15}{\includegraphics{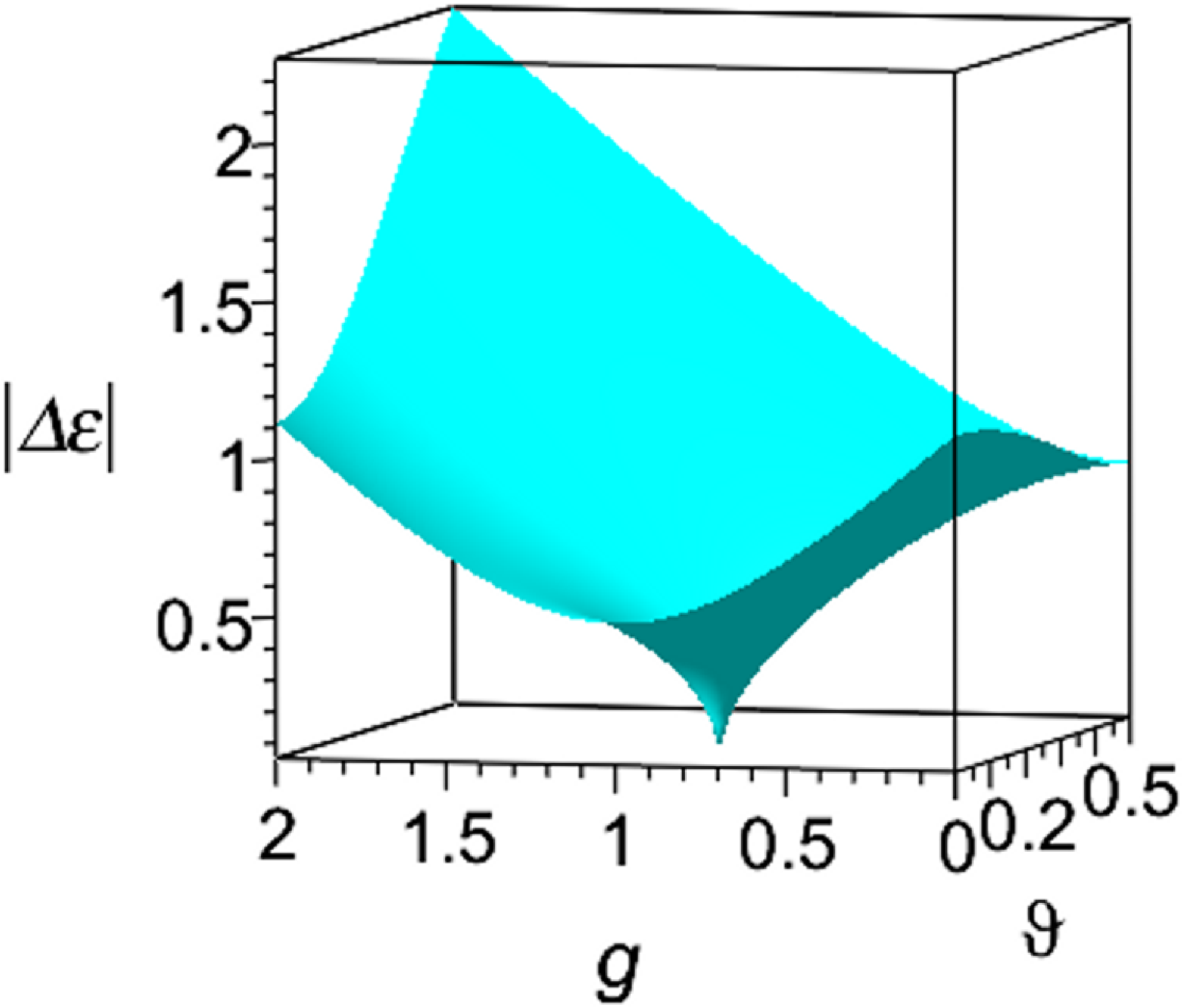}}
\caption{(Color online) Absolute value of difference, $|\Delta \varepsilon|$, of the eigenvalues of the Hamiltonian (\ref{H1a}) as functions of $g$ and $\vartheta$ in the thermodynamic limit. Left panel: $\delta=0$. Right panel: $\delta=0.25$.}
\label{EgV}
\end{figure}

Fig. \ref{EgV} shows in the thermodynamic limit the absolute value of the difference between the two eigenvalues of the Hamiltonian (\ref{H1g}),
\begin{align}\label{EV}
 |\Delta \varepsilon| = 2J|\sqrt{\tilde g^2 -  2\tilde g\cos \vartheta +1}|,
\end{align}
as functions of $g$ and $ \vartheta$. As one can see, the energy gap vanishes at the critical point,
 \begin{align}
& \vartheta_c = \arccos\sqrt{1-\delta^2}, \\
&
g_c=\sqrt{1-\delta^2}.
\end{align}
The difference between the Hermitian QA and non-Hermitian QA is that, while in the first case the gap vanishes for long wavelength modes ($ \vartheta_c =0 $), in the second case the minimal gap shifts to short wavelength modes ($\vartheta_c = \arccos\sqrt{1-\delta^2}$).

\section{ Quench dynamics}

In this section, we consider the NQA for the time-dependent Hamiltonian of Eq.(\ref{EqH1}) written as
\begin{align}\label{EqH2}
\mathcal {\tilde H}_\tau(t)= {\mathcal H}_0 + {\mathcal H}_1(t),
\end{align}
where
\begin{align}
&{\mathcal H}_0 =  -\frac{J}{2}\sum^N_{n=1}\sigma^z_n \sigma^z_{n+1}, \\
&{\mathcal H}_1(t) = -\frac{J}{2}\sum^N_{n=1}\big(g(t) \sigma^x_n + i2\delta(t)\sigma^{-}_n \sigma^{+}_n\big).
\end{align}

We start with the ground state of the auxiliar Hamiltonian, ${\mathcal H}_1(0)$, as the initial state, which is a ``paramagnetic'' with all spins oriented along the $x$ axis. For $g \gg 1$ the Hamiltonian, $\mathcal H_\tau(0)$, is dominated by $\mathcal H_1(0)$, and the ground state of the total Hamiltonian, $\mathcal {\tilde H}_\tau$, is determined by the ground state of  $\mathcal H_1(0)$. The ${\mathcal H}_1$ term causes quantum tunneling between the eigenstates of the Hamiltonian, ${\mathcal H}_0$. At the end of NQA we obtain, $\mathcal {\tilde H}_\tau(\tau)= {\mathcal H}_0 $.  If the quench is slowly enough, the adiabatic theorem guarantees reaching the ground state of the main Hamiltonian, $\mathcal H_0$, at the end of computation.

As shown in the Sec. III, the total Hamiltonian, $\mathcal {\tilde H}_\tau(t)$, in the momentum representation splits into a sum of independent terms, $ \mathcal {\tilde H}_\tau(t)= \sum_{k}\mathcal H_k(t)$. Each $\mathcal H_k$ acts in the two-dimensional Hilbert space spanned by $|k_1\rangle =  |1\rangle_k  |1\rangle_{-k}$ and $ |k_0\rangle= |0\rangle_k  |0\rangle_{-k}$. The wave function can be written as, $|\psi\rangle = \prod_{k}|\psi_k\rangle$, where
\begin{align}\label{Eq10}
|\psi_k(t)\rangle=   c_0(k,t)|k_0\rangle + c_1(k,t) |k_1\rangle.
\end{align}

Choosing the basis as,  $|k_1\rangle =\scriptsize\begin{pmatrix}
                   1 \\
                   0\\
                 \end{pmatrix}
 $ and $|k_0\rangle =\scriptsize\begin{pmatrix}
                   0 \\
                   1\\
                 \end{pmatrix}$, we find that the Hamiltonian, $ \mathcal {H}_k(t)$, projected on this two-dimensional subspace takes the form
\begin{align}\label{H1g}
 \mathcal {H}_k(t)  = -{\varepsilon_0(t)} {1\hspace{-.125cm}1} + {J} \left(
\begin{array}{cc}
              \tilde g(t)- \cos \varphi_k & \sin \varphi_k \\
             \sin \varphi_k & -\tilde g(t)+ \cos \varphi_k  \\
            \end{array}
          \right),
\end{align}
where $\varepsilon_0(t)= J\cos \varphi_k+ iJ\delta(t)$ and $\tilde g(t) = g(t) + i\delta(t)$. Further we assume linear dependence of $\tilde g(t)$ on time:
\begin{align}
\tilde g(t) = \left\{
\begin{array}{l}
\gamma (\tau-t), \quad 0 \leq t \leq \tau \\
0 , \quad t > \tau
\end{array}
\right.
\end{align}
where $\gamma=(g +i\delta)/\tau$, and $g$, $\delta$ are real parameters.

\subsection{Diabatic basis}

The general wave functions,  $|\psi_k\rangle$ and  $\langle\tilde \psi_k |$, satisfy the Schr\"odinger equation and its adjoint equation
 \begin{eqnarray}\label{Eqh1}
i\frac{\partial }{\partial t}|\psi_k\rangle&= \mathcal {H}_k(t)|\psi_k\rangle , \\
-i\frac{\partial }{\partial t}\langle\tilde \psi_k |& =
\langle\tilde \psi_k |\mathcal {H}_k(t)\label{NS2}.
\end{eqnarray}
Presenting $|\psi_k(t)\rangle$ as a linear superposition,
\begin{align}\label{S1}
|\psi_k(t)\rangle = (u_k(t)|k_0\rangle + v_k(t)|k_1\rangle) e^{i\int \varepsilon_0(t)dt},
\end{align}
and inserting expression (\ref{S1}) into Eq. (\ref{Eqh1}), we obtain
\begin{align}\label{IS2a}
i\dot  u_k &= J\big(-(\tilde  g - \cos \varphi_k)\,u_k + \sin \varphi_k\, v_k\big), \\
i\dot  v_k &= J\big(\sin \varphi_k\, u_k +(\tilde  g - \cos \varphi_k)\,v_k \big ).
\label{IS2b}
\end{align}
The solution can be written in terms of  the parabolic cylinder functions, $D_{-i\nu_k}(\pm z)$. (For details see Appendix A.):
\begin{align}\label{Aq4}
& U_{k}(z_k)= B_k D_{-i\nu_k}(z_k)  -{i}{\sqrt{i\nu_k }}  A_k D_{i\nu_k-1}(iz_k)  ,\\
& V_{k}(z_k)=   A_k D_{i\nu_k}(iz_k) -\sqrt{i\nu_k } B_k D_{-i\nu_k-1}(z_k)  ,
\label{Aq4a}
 \end{align}
in which we introduced new functions: $u_k(t)= U(z_k)$, $v_k(t) = V(z_k)$ and
\begin{align}\label{Eq4b}
&z_k(t) = e^{i\pi/4}\sqrt{\frac{2J}{\gamma}}\Big(\gamma(\tau -t)-\cos \varphi_k\Big), \\
&\nu_k= \frac{J\sin^2 \varphi_k}{2\gamma}.
\label{Eq4c}
\end{align}
The constants, $A_k$ and $B_k$ in Eqs. (\ref{Aq4}), (\ref{Aq4a}), are determined by the initial conditions.

In what follows we assume that the evolution of the spin chain starts at $t_0 = 0$ in the `ground' state. This implies that for each $k$ the evolution of the corresponding two-level system (TLS) starts from the state, $|\psi(0)\rangle = |k_0\rangle $. Then, the following initial conditions should be imposed: $u_k(0)=1$ and $v_k(0)=0$. The related boundary condition are,  $z_k(0) = e^{i\pi/4}\sqrt{2J/\gamma}\big(\gamma\tau -\cos \varphi_k\big)$. Using these conditions, we obtain the solution of the Schr\"odinger equation as follows. (See Appendix A for details.):
\begin{align}\label{eq5a}
& U_{k}(z_k)= B_k D_{-i\nu_k}(z_k),\\
& V_{k}(z_k)=  -\sqrt{i\nu_k } B_k D_{-i\nu_k-1}(z_k),
\label{eq5b}
 \end{align}
where $B_k= e^{\pi\nu_k/2}  D_{i\nu_k}\big(iz_k(0)\big)$.

It is assumed that a quantum measurement will determine the state of the quantum system at $t > \tau$, when the external field, $\tilde g(t)=0$. We denote the final state of the system, at $t=\tau$, as $|\psi_\tau\rangle = \prod_{k} |\psi_k(\tau)\rangle $. The probability, $P_n(k)$, of finding the TLS in a given state, $|k_n\rangle $ $(n=0,1)$, can be written as
\begin{align}\label{Eq9}
 P_n(k) = \frac{|\langle k_n|\psi_\tau\rangle|^2}{ |\langle \psi_\tau|\psi_\tau\rangle|^2}.
\end{align}

Since for non-Hermitian systems the norm of the wavefunction is not conserved, we define the (intrinsic) probability of transition, $|k_0\rangle$ $\rightarrow$ $|k_1\rangle$, as
\begin{align}\label{QEq2}
 P_k(t)= \frac{|v_k(t)|^2}{|u_k(t)|^2 + |v_k(t)|^2}.
\end{align}
Using the functions, $U(z_k)$ and $V(z_k)$, we recast (\ref{QEq2}) as
\begin{align}\label{Eq2a}
 P_k(t)= \frac{1}{1+ \displaystyle\frac{| D_{-i\nu_k}(z_k(t)) |^2}{|\sqrt{i\nu_k}\, D_{-i\nu_k-1}(z_k(t)) |^2}}.
\end{align}

To calculate $P_k(t)$ at the end of evolution ($t=\tau$) we use asymptotic formulas of the Weber functions. For large values of the argument, $|z_k(\tau)|=|\sqrt{2J/\gamma}\cos \varphi_k| \gg 1$, one can apply the asymptotic formulas for parabolic cylinder functions to estimate the probability of transition. For $\tau \gg 1$ the modulus of this argument is large for most $k$, except near $\varphi_k  = \pm \pi/2$.

For wavelength modes with $\varphi_k  \ll \pi/4$, using the asymptotic formulas for the Weber functions, we obtain
\begin{align}\label{Eq5}
\frac{ D_{-i\nu_k}(z_k(\tau)) }{\sqrt{i\nu_k}\, D_{-i\nu_k-1}(z_k(\tau))}\approx \frac{\displaystyle e^{-\pi \nu_k/2}e^{-z_k^2(\tau)/2}\Gamma(1+i\nu_k )}{\sqrt{2\pi i\nu_k }}.
\end{align}
Inserting (\ref{Eq5}) into Eq. (\ref{Eq2a}), we obtain
\begin{align}\label{PA1}
P_k(\tau) = \frac{1}{1+ \displaystyle\frac{|\Gamma(1+i\nu_k )|^2 }{2\pi |\nu_k| }\,e^{-\pi\Re\nu_k- \Re z^2_k(\tau)}}.
\end{align}

For $\delta \ll g$ we can approximate
$\Gamma(1+i\nu )\approx \Gamma(1+i\Re\nu )$. Next, using the relation \cite{Leb},
\begin{align}
|\Gamma(iy)|^2 = \frac{\pi}{y \sinh \pi y},
\end{align}
for real $y$, we obtain
\begin{align}\label{Pq2k}
P_k(\tau) =\frac{1-  e^{-2\pi\Re\nu_k}}{1-e^{-2\pi\Re\nu_k}+  e^{-2\pi\Re\nu_k- \Re z^2_K(\tau)}}.
\end{align}

In the case of the Hermitian QA ($\delta =0$),one has $\Re z^2_k(\tau)=0$, and Eq. (\ref{PA1}) leads to the Landau-Zener formula \cite{LL,ZC},
\begin{eqnarray}\label{P1}
P_k(\tau) =1-e^{-(\pi J\tau/g)\sin^2 \varphi_k}.
\end{eqnarray}

To calculate the transition probability for  $\varphi_k \approx \pi/2$ we use the expansion for small value of the argument of Weber function, $|z_k | \ll 1$ (see Eq. (\ref{W0})).
The computation yields
\begin{align}\label{Eq8}
P_{\pi/2}(\tau) =  \frac{1}{ 1+ \displaystyle\frac{|\Gamma (\frac{i\nu_k}{2})|^2}{|\Gamma (\frac{1+i\nu_k}{2})|^2}}.
\end{align}
For the Hermitian QA this gives
\begin{align}\label{Eq8a}
P_{\pi/2}(\tau) = \frac{\tanh\pi\nu/2}{1+ \tanh\pi\nu/2},
\end{align}
in which $\nu = J\tau/(2g)$. From Eq. (\ref{Eq8a}) it follows that for $\tau \gg g/J$ the probability $P_{\pi/2}(\tau) \simeq 1/2$.

\begin{figure}[tbp]
\scalebox{0.215}{\includegraphics{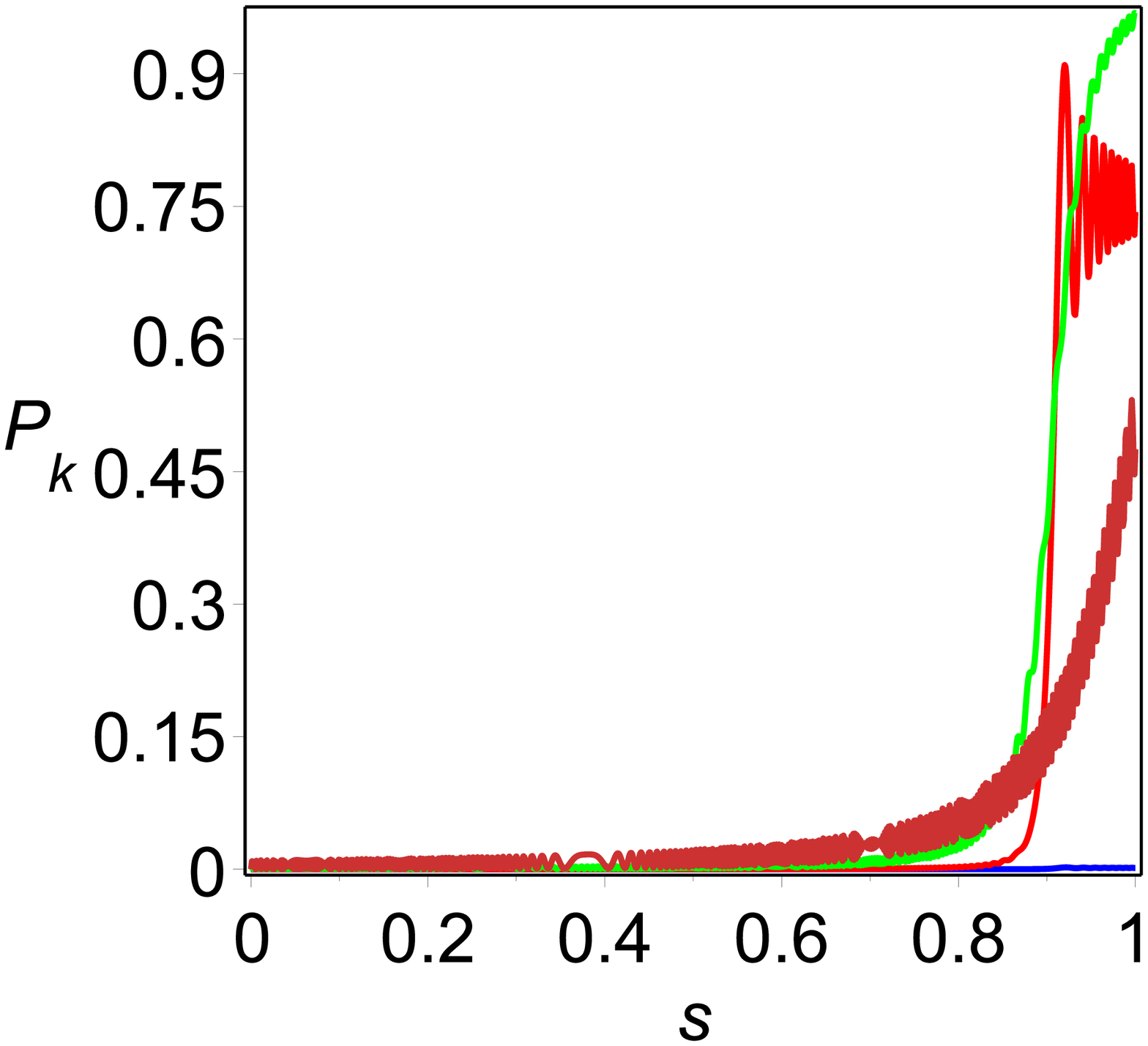}}
\scalebox{0.21}{\includegraphics{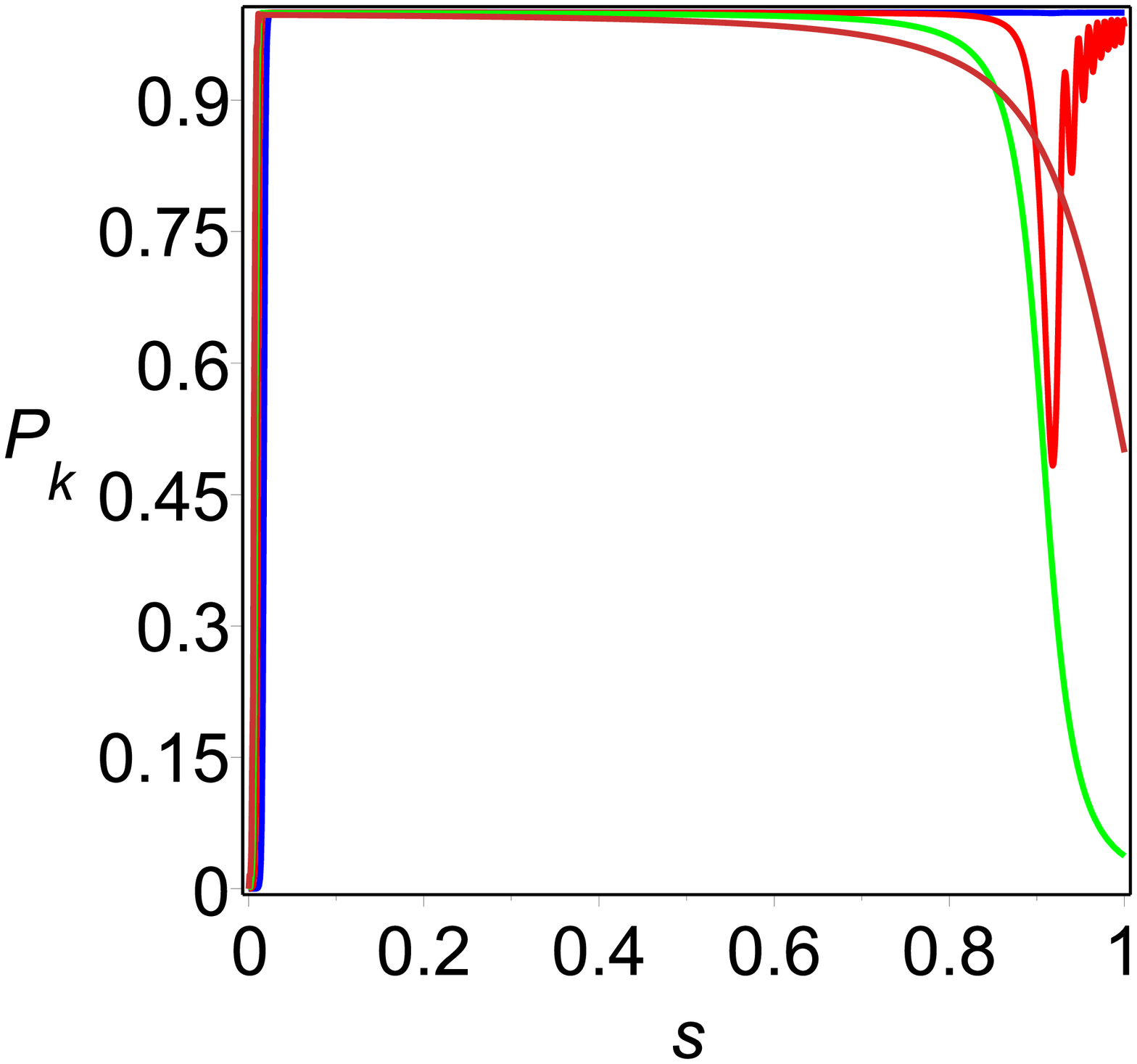}}
\caption{The probability of transition (remain in the ground state), $|k_0\rangle$ $\rightarrow$ $|k_1\rangle $, as a function of the scaled time, $s =t/\tau $ ($J=0.5$, $g=10$, $\tau = 10^3$, $N=1024$, $k=p-1/2$). Left panel: Hermitian QA ($\delta =0$). From bottom to top: $p=1,256,16,64$. Right panel: NQA ($\delta = 0.5$). From top to bottom: $p=1,16,256,64$. }
\label{P1a}
\end{figure}

In Fig. \ref{P1a} we present the results of our numerical simulation for $N=1024$ qubits. As one can see, for long wavelength modes with $\varphi_k  \ll\pi/4$ (blue and red lines), the NQA shows better performance than the Hermitian QA. For $\varphi_k =\pi/2$ the probability of transition is $P_{\pi/2}(\tau) \simeq 1/2$ for both schedules, either Hermitian QA or non-Hermitian QA (orange line).

\subsection{Adiabatic basis}

A widely used opinion is that for slowly enough evolution the only long wavelength modes become excited (see e.g. \cite{DJ}). However, as it was shown in the Subsection A, even for the Hermitian QA, the transition probability, $P_k \approx 1/2$, for $k\approx \pi/2$. Thus, it is not clear what is the contribution of the short wavelength modes ($ k \sim \pi/2$) to the probability of the whole system to stay in the ground state. To respond to this question we consider the expansion of the wavefunction, $|\psi\rangle = \prod_k |\psi_k(t) \rangle $, in the adiabatic basis, formed by the instantaneous eigenvectors.

In the adiabatic basis the wavefunction, $|\psi_k(t) \rangle$, can be written as follows:
\begin{align}\label{Eq7}
 |\psi_k(t) \rangle = \alpha_k(t) |u_{-}(k,t)\rangle + \beta_k(t) |u_{+}(k,t)\rangle.
\end{align}
We assume that the evolution begins from the ground state that implies, $\alpha_k(0)=1$ and $\beta_k(0)=0$.
At the end of evolution at $t=\tau$, when $\tilde g=0$, we have
\begin{align}\label{E3}
      |\psi_k(\tau) \rangle= \alpha_k(\tau) |u_{-}(k,\tau)\rangle + \beta_k(\tau) |u_{+}(k,\tau)\rangle,
\end{align}
where
\begin{align}\label{Eq1}
&|u_{+}(k,\tau)\rangle = \left (\begin{array}{c}
                \sin\frac{\varphi_k}{2}  \\
             \cos \frac{\varphi_k}{2}\\
            \end{array}
          \right), \\
&|u_{-}(k,\tau)\rangle = \left (\begin{array}{c}
               - \cos\frac{\varphi_k}{2}  \\
             \sin \frac{\varphi_k}{2}\\
            \end{array}
          \right).
\end{align}

By presenting,
\begin{align}\label{Eq5f}
 \alpha_k(t) = a_k(t)\,e^{-i\int_0^t \varepsilon_{-}(k,t) dt}, \\
  \beta_k(t) = b_k(t)\,e^{-i\int_0^t \varepsilon_{+}(k,t) dt},
  \label{Eq5g}
\end{align}
one can show that the coefficients, $a_k(t)$ and $b_k(t)$, satisfy the following asymptotic conditions:
\begin{eqnarray}\label{V2}
a_k(\tau) = 1 + {\mathcal O}(1/\tau), \\
b_k(\tau) =  {\mathcal O}\bigg( \exp{\Big(2\tau\Im{ \int_0^{z_c} \varepsilon_k(z) dz}\Big)}\bigg),
\label{V2a}
\end{eqnarray}
where the critical point, $z_c$, lies on the first Stokes line in the {\em lower} complex line defined as
\begin{equation}\label{St1}
   \Im{ \int_0^{z_c} \varepsilon_k(z) dz} < 0.
\end{equation}
Here the critical point, $z_c$, is determined as a solution of the equation, $\varepsilon_k(z_c)=0$, in the complex plane obtained by analytical continuation, $t \rightarrow z$ \cite{BMV1,JMPC,JKP,KP,SVG,DG, DAM,DJP,HJP}.

Similarly, if initially the system was in the excited state: $|\psi_k(0)\rangle = |u_{+}(k,0)\rangle$, so that $\alpha_{k}(0) =0$ and  $\beta_{k}(0) =1$, the result of integration yields
\begin{eqnarray}\label{V3}
b_k(\tau)  = 1 + {\mathcal O}(1/\tau), \\
 a_k(\tau) =  {\mathcal O}\bigg(\exp{\Big(2\tau\Im{ \int_0^{z_c} \varepsilon_k(z) dz}\Big)}\bigg).
\label{V3a}
\end{eqnarray}

The intrinsic probability to remain in the ground state at the end of the adiabatic evolution is given by
\begin{align}\label{Eq2}
P^{gs}_k(\tau) = \frac{|\alpha_k(\tau)|^2}{|\alpha_k(\tau)|^2 + |\beta_k(\tau)|^2}.
\end{align}
With help of Eqs. (\ref{Eq5f}), (\ref{Eq5g}) we obtain
\begin{align}\label{AEq1}
P^{gs}_k(\tau) = \frac{1}{1 + \displaystyle \frac{|b_k(\tau)|^2}{|a_k(\tau)|^2}\,e^{4\Im \int_0^\tau \varepsilon_k(t)dt}}.
\end{align}
From here it follows that for any, $0 \leq t\leq \tau$, the adiabatic evolution should be performed along the path corresponding to,  $\Im \varepsilon_k(t) \leq 0$ .
In the exact solution, given by
\begin{align}\label{eq5g}
\alpha_k(\tau) =&- B_k e^{-\delta\tau/2}\bigg( D_{-i\nu_k}(z_k(\tau))\sin \frac{\varphi_k}{2} \nonumber \\
& + \sqrt{i\nu_k }  D_{-i\nu_k-1}(z_k(\tau))\cos\frac{\varphi_k}{2}    \bigg) ,\\
\beta_k(\tau) =& B_k e^{-\delta\tau/2}\bigg( D_{-i\nu_k}(z_k(\tau))\cos\frac{\varphi_k}{2} \nonumber \\
&- \sqrt{i\nu_k }  D_{-i\nu_k-1}(z_k(\tau))\sin\frac{\varphi_k}{2}    \bigg),
\label{eq5e}
 \end{align}
it manifests itself in the choice of phase in the argument of the Weber function, when we apply the asymptotic expansion.

The probability for the whole system to stay in the ground state at the end of the evolution is given by, $P_{gs}= \prod_k P^{gs}_k(\tau)$. For long wavelength modes with $\varphi_k\ll \pi/4$, using the asymptotic formulas for the Weber functions with the large value of its argument, we find that $P^{gs}_k(\tau) = P_k(\tau)$, where $P_k(\tau)$ is the transition probability of spin flip given by Eq. (\ref{PA1}). For the Hermitian QA this yields the LZ result
\begin{align}\label{Pq1}
P_k(\tau) =1-e^{-(\pi J\tau/g )\sin^2 \varphi_k} \approx 1-e^{-\pi^3 J\tau k^2/(gN^2)}.
\end{align}
And in the case of the NQA (for $\delta \ll g$) we obtain
\begin{align}\label{Pq2}
P_k(\tau) =\frac{1-  e^{-2\pi\Re\nu_k}}{1-e^{-2\pi\Re\nu_k}+  e^{-2\pi\Re\nu_k- \Re z^2_k(\tau)}},
\end{align}
where
\begin{align}\label{Eq3a}
 &\Re\nu_k \approx (J\tau /2g) \sin^2 \varphi_k,  \\
& \Re z^2_k(\tau) \approx (2\delta J\tau/g^2) \cos^2 \varphi_k .
 \label{Eq3b}
\end{align}

For short wavelength modes, approximately with $\pi/4 < \varphi_k \leq \pi/2$, employing the large-order asymptotics for Weber functions, we obtain $P^{gs}_k(\tau) = 1 + {\mathcal O}(1/\sqrt{|\nu_k|})$. (See Appendix A.)

\begin{figure}[tbp]\label{QA_1}
\begin{center}
\scalebox{0.3}{\includegraphics{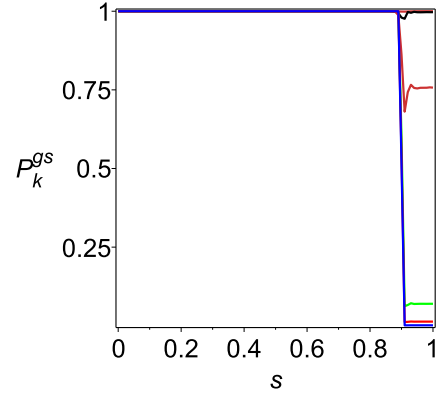}}
\end{center}
\caption{(Color online) The probability, $P^{gs}_k$, of TLS to remain in the ground state as a function of the scaled time, $s =t/\tau $, for the Hermitian QA ($\delta = 0$, $J=0.5$, $g=10$, $\tau = 10^3$, $N=1024$, $k =p-1/2$). From bottom to top: $ p=1,2,4,16,32,64$. }
\label{P2a}
\end{figure}

\begin{figure}[tbp]
\scalebox{0.3}{\includegraphics{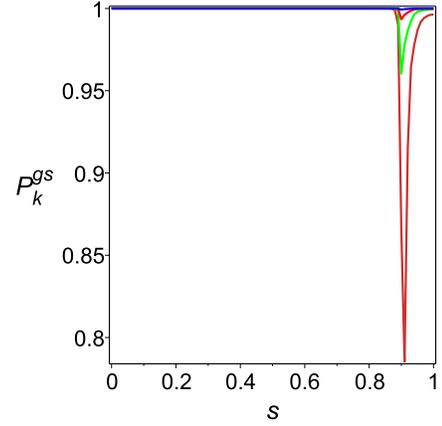}}
\caption{(Color online) The probability, $P^{gs}_k$, of TLS to remain in the ground state as a function of the scaled time, $s =t/\tau $, for the NQA ($\delta = 0.5$, $J=0.5$, $g=10$, $\tau = 10^3$, $N=1024$, $k =p-1/2$ ). From bottom to top: $ p=1,2,4,16$.
 }
\label{P2b}
\end{figure}

Our theoretical predictions are confirmed by numerical calculations performed for $N=1024$ qubits. (See Figs. \ref{P2a}, \ref{P2b}.) One can observe that while short wavelength excitations are essential at the critical point, at the end of evolution their contribution to the transition probability from the ground state to the first excited state is negligible.

\section{Performance characterization of the quantum annealing}

During the QA the system does not stay always at the ground state at all times. At the critical point, the system becomes excited, and its final state is determined by the number of defects (kinks). To evaluate the efficiency of QA  one can calculate the number of defects. Then, computational time is the time required to achieve the number of defects below some acceptable value.

Following \cite{DJ}, we define the operator of the number of kinks as
\begin{align}
 \hat{\mathcal N} = \frac{1}{2}\sum^N_{n=1}\big ( 1- \sigma^z_n \sigma^z_{n+1}\big ) = \sum_{k>0} (b_k^\dagger b_k+ b_{-k}^\dagger b_{-k}).
\end{align}
The number of kinks is equal to the number of quasiparticles
excited at $\tilde g=0$ (final state). It is given by
\begin{align} \label{K2}
{\mathcal N} =  \langle \psi_\tau |\hat{\mathcal N} | \psi_\tau \rangle = \frac{1}{J}\sum_{k>0} \big(\langle \psi_k(\tau) |{\mathcal H}_k(\tau) | \psi_\tau \rangle - E_{gs} \big).
\end{align}
This can be recast as follows: ${\mathcal N} = E_{res}/J$, where $E_{res}$ is the residual energy defined as the difference between the solution obtained by QA and the exact one \cite{SMTC,SSMO,SSMOK},
\begin{align}
E_{res}= \sum_{k>0} \big(\langle \psi_k(\tau) |{\mathcal H}_k(\tau) | \psi_\tau \rangle - E_{gs} \big) = J\mathcal N.
\end{align}
Thus, an alternative (equivalent) way to evaluate the efficiency of QA is to calculate $E_{res}$. Note, that for more complicated systems that include disorder, the residual energy may not have a simple link with the density of defects \cite{CFS}.

Using Eq. (\ref{E3}), we can calculate the number of kinks as
\begin{align}\label{TL1}
\mathcal N = \sum_{k>0}( 1- P^{gs}_k(\tau)),
\end{align}
where $P^{gs}_k(\tau)$ is given by Eq. (\ref{Eq2}). In the thermodynamic limit the sum in Eq. (\ref{TL1}) can be replaced by integral, and we obtain for the density of kinks the following expression:
\begin{align}\label{QP1a}
n= \lim_{N\rightarrow \infty} \frac{\mathcal N}{N} = \frac{1}{\pi}\int^{\pi}_{0} d \vartheta( 1- P^{gs}(\vartheta,\tau)).
\end{align}

As was shown in the previous section, during slow evolution only long wavelength modes can be excited. So one can use the Gaussian distribution by replacing $\sin \varphi_k \approx \varphi_k $ and $\cos \varphi_k \approx 1$. In the limit $\sqrt{2J\tau/g} \gg 1$, we can employ Eqs. (\ref{Pq2}) - (\ref{Eq3b}) to calculate the number of kinks as
\begin{align}\label{QP1}
n=  \frac{1}{\pi}\int^{\pi}_{0}  \frac{e^{-2\pi\Re\nu- \Re z^2}d \vartheta}{1-e^{-2\pi\Re\nu}+  e^{-2\pi\Re\nu \Re z^2}}.
\end{align}

Performing the integration with $\Re\nu = J\tau \vartheta^2 /(2g)$  and $ \Re z^2= 2\delta J\tau/g^2$, we obtain \footnote{Approximation $\cos \varphi_k \approx 1$ instead of $\cos \varphi_k \approx 1- \varphi_k ^2/2$ leads to omission of corrections to the formula (\ref{Eq10}) up to $\mathcal O(\delta/g)$.}
\begin{align}\label{Eq10h}
n =  n_0 e^{-2\delta\tau J/g^2}\Phi\Big( 1-e^{-2\delta\tau J/g^2},\frac{1}{2},1\Big),
\end{align}
where
\begin{align}\label{n1}
 n_0 = \frac{1}{2\pi}\sqrt{\frac{g}{J\tau}}
\end{align}
denotes the density of kinks for the Hermitian LZ problem \cite{DJ}, and  $\Phi(x,a,c)$ is the Lerch transcendent \cite{EMOT1}.

\begin{figure}[tbp]
\scalebox{0.3}{\includegraphics{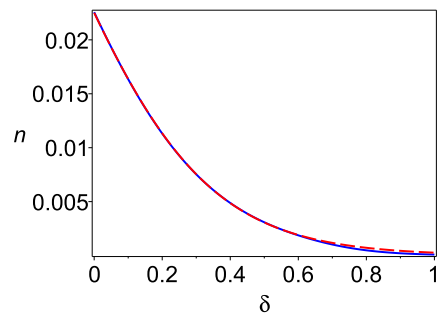}}
\caption{(Color online) Density of kinks as function of the dissipation parameter, $\delta$ ($\tau = 10^3$, $J=0.5$, $g=10$, $N=1024$). Blue line: exact result. Red dashed line: asymptotic formula of Eq. (\ref{Eq10h}).}
\label{P2ka}
\end{figure}

\begin{figure}[tbp]
\scalebox{0.225}{\includegraphics{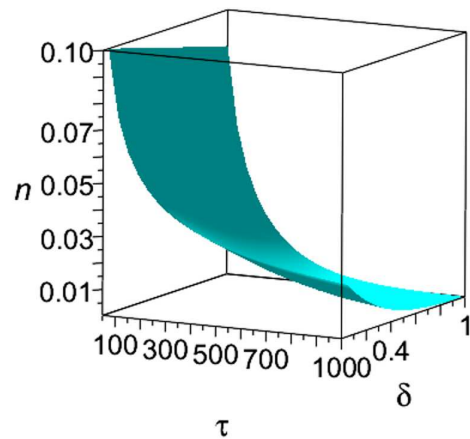}}
\caption{(Color online) Density of kinks as function of the dissipation parameter $\delta$ and annealing time $\tau$.}
\label{P2h}
\end{figure}

In Figs. \ref{P2ka}, \ref{P2h} we present the results of numerical simulation for the density of defects. In Fig. \ref{P2ka} the density of kinks as a function of $\delta$ is depicted. In Fig. \ref{P2h} we show the dependence of the density of defects as a function of the decay parameter, $\delta$, and annealing time, $\tau$, in the thermodynamic limit. This is consistent with the results of numerical simulation presented in Ref. \cite{ZDZ} for the Hermitian LZ problem.

\begin{figure}[tbp]
\scalebox{0.35}{\includegraphics{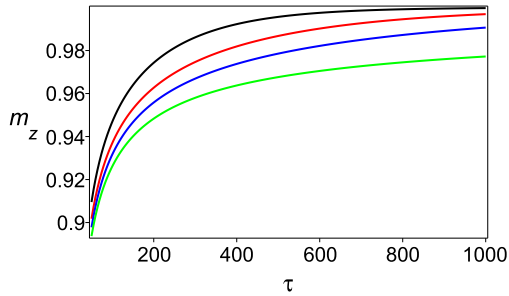}}
\caption{(Color online) Density of magnetization, $m_z$, as function of the annealing time, $\tau$. From bottom to top: $\delta=0,0.25,0.5,1$ ($J=0.5$, $g=10$).}
\label{mz}
\end{figure}

The final state of the system is a ferromagnetic state with the finite domains of spins (pointed up or down), separated by kinks. The magnetization, $Nm_z$, is defined from the expression for the total energy of the ground state: $E_{gs}=-NJm_z^2$. In the thermodynamic limit, we obtain
\begin{align}\label{mz1}
m_z= \lim_{N\rightarrow \infty}\sqrt{\frac{|E_{gs}|}{JN}} = \sqrt{1-2n}.
\end{align}
In Fig. \ref{mz} the density of magnetization as function of the annealing time, $\tau$,
is depicted. As one can see, even moderate dissipation essentially decreases the number of defects in the system.
\begin{figure}[tbp]
\scalebox{0.3}{\includegraphics{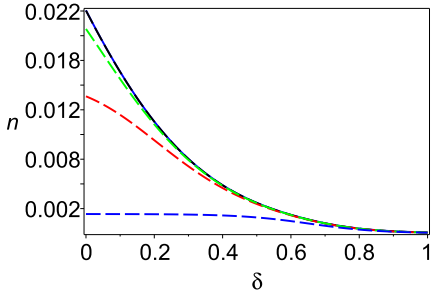}}
\caption{(Color online) Density of kinks as function of the dissipation parameter, $\delta$ ($J=0.5$, $g=10$, $\tau = 10^3$, $N=1024$). Blue solid shows the exact result. Dashed color lines present contribution of the first $k$-modes ($k=p-1/2$). From bottom to top: $p=1,8,16,32$ . }
\label{P2k}
\end{figure}

Due to the symmetry of the problem with respect to $k \rightarrow -k$, and since for each $k$  there is independent evolution, the probability of the whole system to remain in the ground state at the end of evolution is the product
\begin{align} \label{Eq4}
P_{gs} = \prod_{k>0} P^{gs}_k(\tau).
\end{align}

In the long wavelength approximation one can take into account only $\varphi_k= \pi/N$, and estimate $ P_{gs}$ as
\begin{align} \label{Eq4a}
P_{gs} \approx \frac{1-  e^{-2\pi\Re\nu}}{1-e^{-2\pi\Re\nu}+  e^{-2\pi\Re\nu- \Re z^2(\tau)}},
\end{align}
where $\Re z^2(\tau) = 2\delta J\tau/g^2$ and $\Re\nu = \tau /\tau_0 $. Here we denote  $\tau_0 = 2gN^2/(\pi^2J)$.

\begin{figure}[tbp]
\scalebox{0.215}{\includegraphics{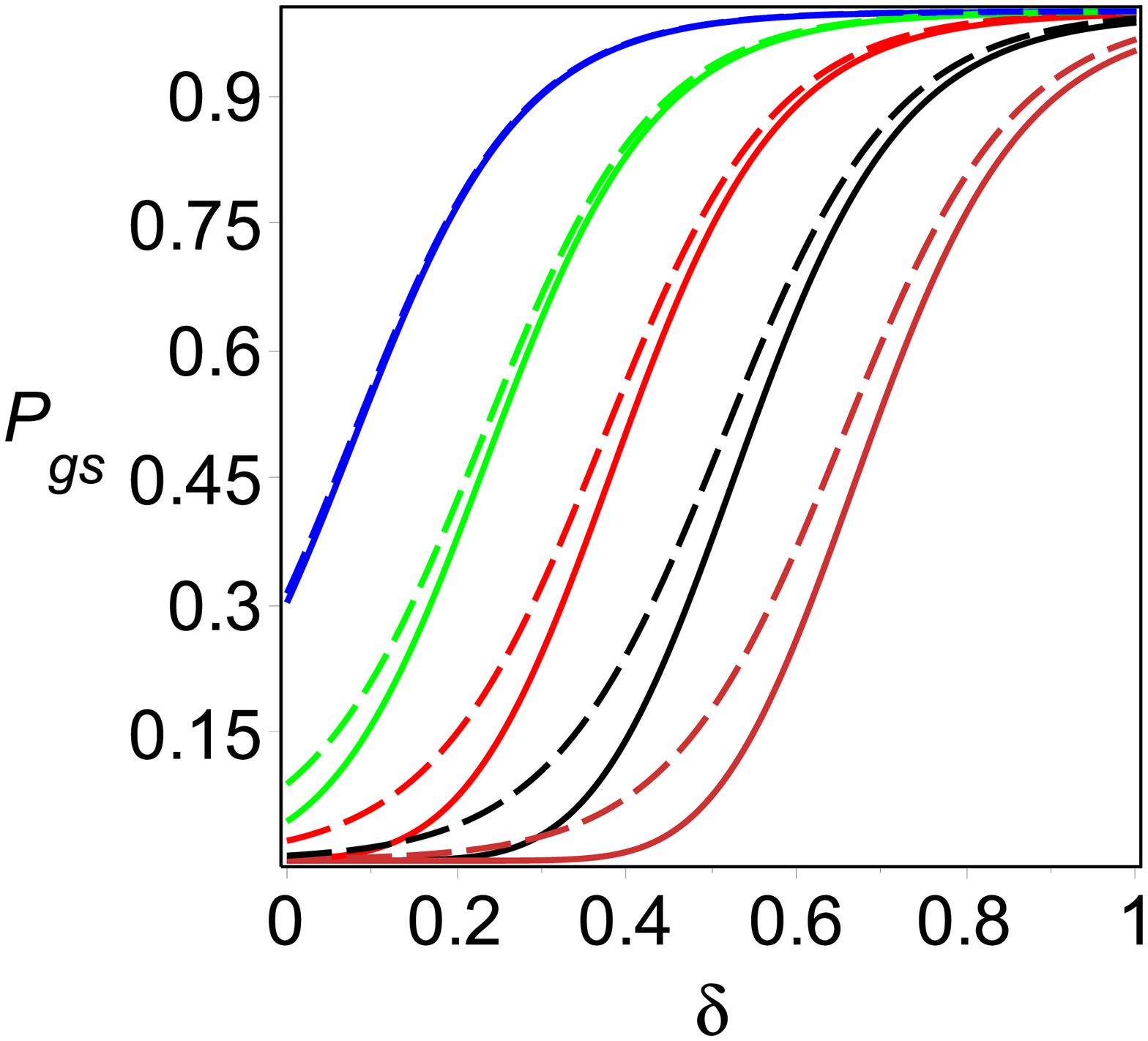}}
\scalebox{0.22}{\includegraphics{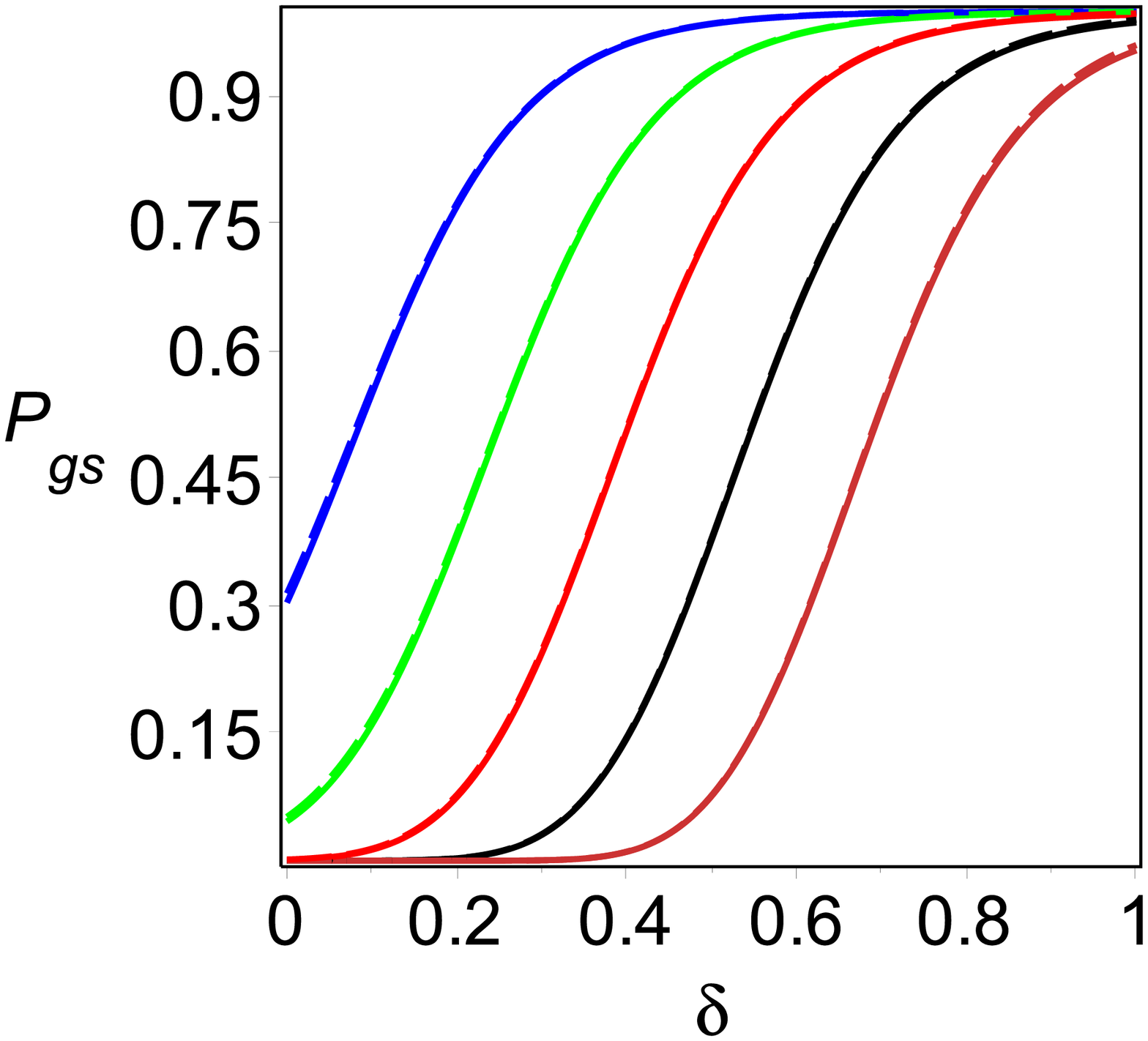}}
\caption{(Color online) Probability to stay in the ground state, $P_{gs}$, as function of the dissipation parameter, $\delta$ ($J=0.5$, $g=10$, $\tau = 10^3$, $k=p-1/2$). Left panel: Solid curves present the exact result. Dashed lines present the contribution of the first mode ($p = 1$). Right panel: Solid curves are the contribution of  all modes. Dashed lines are the contribution of the first $p=1,2, \dots, N/64$ modes. From top to bottom: $N=64,128,256,512,1024$. }
\label{Pgs_3}
\end{figure}

\begin{figure}[tbp]
\scalebox{0.225}{\includegraphics{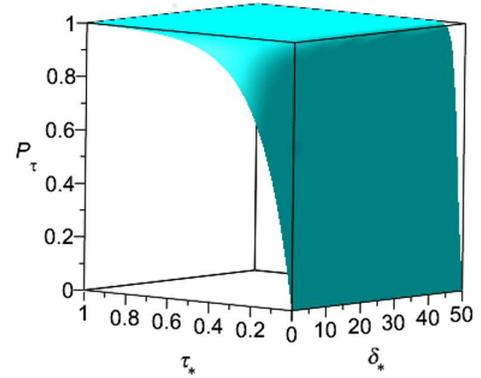}}
\caption{(Color online) The transition probability, $P_\tau$, as function of a scaled decay rate, $\delta_\ast=\delta \tau_0/g^2$, and scaled annealing time, $\tau_\ast =\tau/\tau_0 $.}
\label{P3d}
\end{figure}

In Figs. \ref{Pgs_3} and \ref{P3d}, the results of numerical simulation are demonstrated. As one can see, for the probability $P_{gs}\approx 1$ the asymptotic formula of Eq. (\ref{Eq4a}) is in a good agreement with the exact formula (\ref{Eq4}). We also performed numerical simulations to demonstrate that for any $N$ the contribution of the first $N/64$ modes yields essentially the same result as the exact formula (\ref{Eq4}). (See Fig. \ref{Pgs_3}.) We find that even moderate dissipation boosts the transition probability.

For the Hermitian QA $(\delta =0)$, Eq. (\ref{Eq4a}) yields the Landau-Zener formula \cite{LL,ZC}
\begin{eqnarray}\label{P}
P_{gs} =1-e^{-2\pi\tau/\tau_0}.
\end{eqnarray}
From here it follows that $P_{gs} \approx 1$, if $\tau \geq \tau_0 $. Thus, the computational time for the Hermitian QA should be of order $N^2$.

For the NQA, assuming $\tau \ll \tau_0$, we obtain
\begin{align}\label{Pq2g}
P_{gs} \approx \frac{1}{1+ \displaystyle\frac{\tau_0}{2\pi\tau}\,  e^{- 2J\delta\tau/g^2}}.
\end{align}
From here, in the limit of $\delta \rightarrow 0$, we obtain
\begin{align}\label{Pq2_k}
P_{gs} \rightarrow \frac{1}{1+ \displaystyle\frac{\tau_0}{2\pi\tau}} \ll 1.
\end{align}
The obtained result is expected, as in this case, the time of the Hermitian annealing, $\tau$, is small with respect to the characteristic time, $\tau_0$: $\tau\ll \tau_0$.

Next, assuming
\begin{align}\label{AT1}
\frac {2J \delta\tau}{g^2} - \ln \frac{\tau_0}{2\pi\tau} \gg 1.
\end{align}
we obtain
\begin{align}\label{Pq2_g}
P_{gs} \approx 1- \frac{\tau_0}{2\pi\tau}\,  e^{- 2J\delta\tau/g^2}.
\end{align}
As one can see, $P_{gs} \approx 1$, if the conditions of Eq. (\ref{AT1}) are satisfied.
From (\ref{AT1}) we obtain the following rough estimate of the computational time for NQA: $\tau \approx (g^2/2J\delta)\ln N$.

\begin{figure}[tbp]
\scalebox{0.275}{\includegraphics{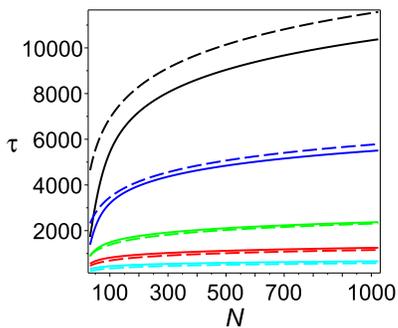}}
\caption{(Color online) Annealing time, $\tau$, as a function of $N$ for NQA ($J=0.5$, $g=10$). Solid lines present the exact result. Dashed lines are the asymptotic formula of Eq.(\ref{T1}). From top to bottom: $\delta=0.05,0.1,0.25,0.5,1 $.}
\label{tau_1}
\end{figure}

To find the annealing time from the exact formula (\ref{Eq4a}) we impose the following condition on the probability: $ P_{\tau}=0.999$. Then, we solved numerically Eq. (\ref{Eq4a}). The results of numerical calculations are presented in Fig. \ref{tau_1}. We find that the best fit of the asymptotic expression to the exact result is given by the following asymptotic formula (see Fig. \ref{tau_1}):
\begin{align}\label{T1}
  \tau \approx \frac{g^2}{2J\delta}\ln\frac{N}{\pi}.
\end{align}
\begin{figure}[tbp]
\scalebox{0.21}{\includegraphics{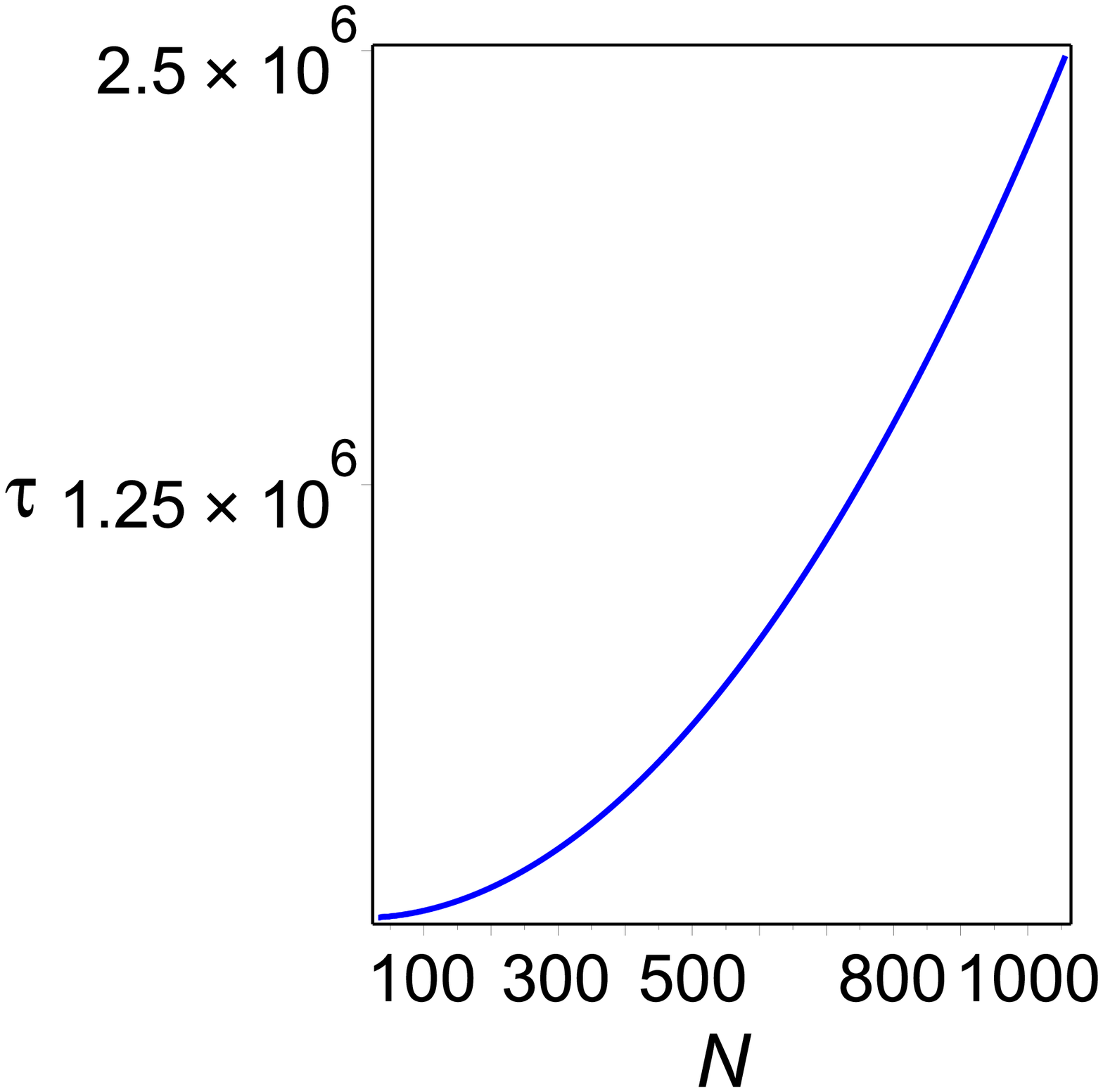}}
\scalebox{0.2}{\includegraphics{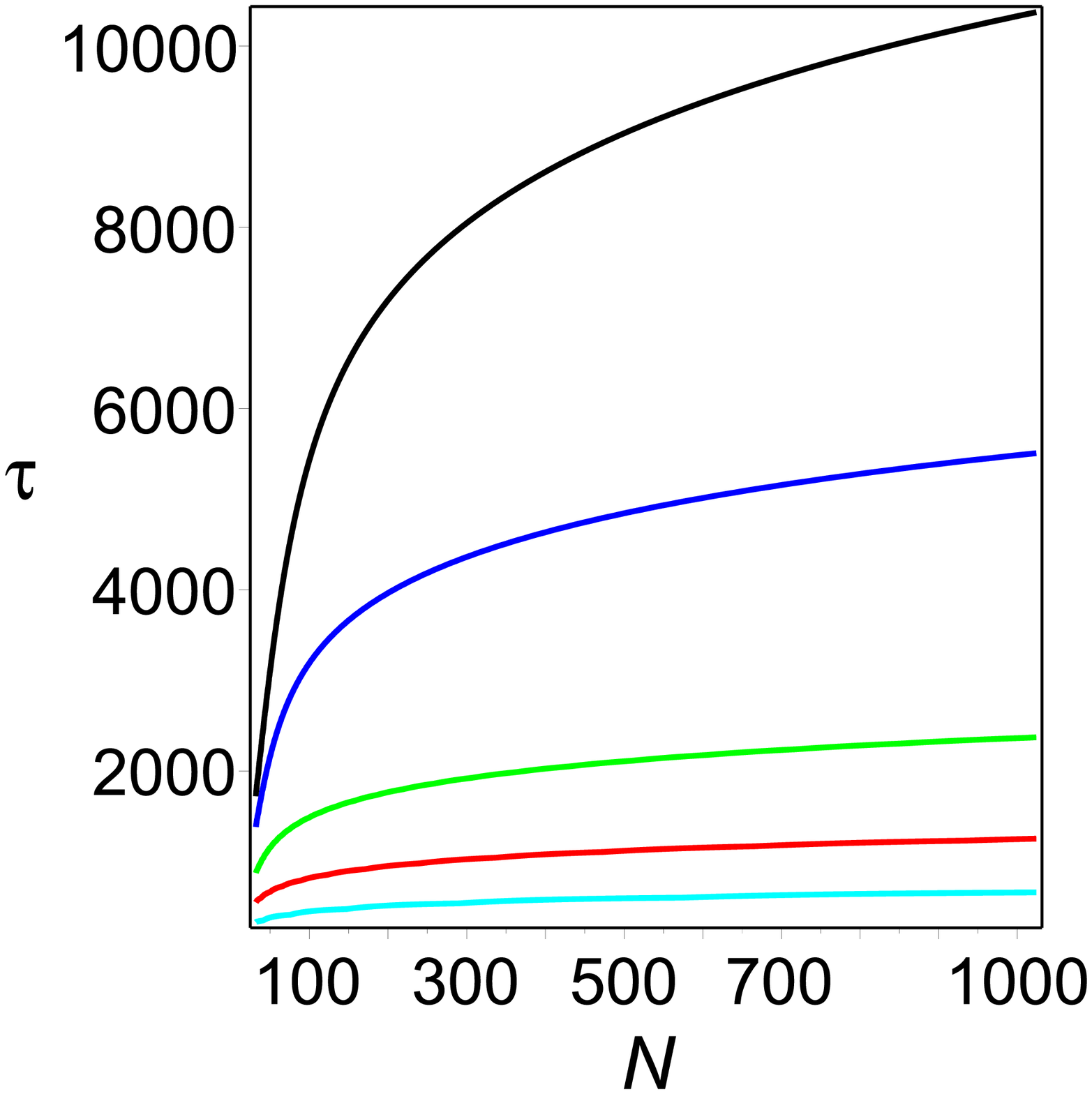}}
\caption{(Color online) Annealing time, $\tau$, as function of $N$. Left panel.  Hermitian QA, $\delta=0$  ($J=0.5$, $g=10$). Right panel. Non-Hermitian QA, from top to bottom: $\delta=0.05,0.1,0.25,0.5,1 $.}
\label{tau}
\end{figure}

In Fig. \ref{tau}, the annealing time, $\tau$, as a function of the number of spins is depicted for NQA (left panel) and Hermitian QA (right panel). The comparison shows that for large number of spins ($N\sim 1000$) and for $\delta\gtrsim 0.25$, the annealing time of NQA is $\approx 10^{3}$ times smaller than for Hermitian QA.

The obtained results indicate that the characteristic time of non-Hermitian annealing, even for small but finite $\delta\not=0$, is defined not only by the number of spins, $N$ (as in Hermitian annealing), but mainly by the dissipation  rate, $\delta$. (See Figs. \ref{Pgs_3}, \ref{P3d} and Eq. (\ref{Pq2g}).) Thus, the non-Hermitian quantum annealing has complexity of order $\ln N$, which is much better than the quantum Hermitian (global) adiabatic algorithm. Also, this complexity is certainly better  than one of the adiabatic local annealing algorithm which has a total running time of order $N$ \cite{RN}.

\section{Conclusions}

Recently, many modifications of quantum annealing algorithms have been proposed \cite{KN,FGGLL,SSMO,DC,SMTC,SST,SNS,AM,OMNH,MO,ON}. The main objective of these publications is to significantly decrease the time of annealing, so that the solutions of hard optimization problems could be obtained either by (i) combining classical computers with quantum algorithms or by (ii) building  real quantum computers. One of the very popular test models is the Ising spin system which is also useful for practical purposes. In this case, the quantum annealing algorithms are used to find the ground state of this system. The approach presented in this paper, is related to item (i) above, in application to the ferromagnetic Ising spin chain. We have chosen an auxiliary Hamiltonian in such a way that the total Hamiltonian is non-Hermitian. This allowed us to shift the minimal gap in the energy spectrum in the complex plane, and significantly reduce the time required to find the ground state. Our approach leads to the annealing time $\sim \ln N$, where $N$ is the number of spins, which is much less than the time of Hermitian annealing ($\sim N^2$) for the same problem. But many serious problems still remain to be considered.  One of them is the application of this dissipative approach to more complicated Ising-type models with frustrated interactions, when the ground state can be strongly inhomogeneous. Another direction is to use both dissipation and pumping into the system, as it was done in \cite{SU,KT}. This research is now in progress.



\acknowledgements

The work by G.P.B. was carried out under the auspices of the National Nuclear Security Administration of the U.S. Department of Energy at Los Alamos National Laboratory under Contract No. DE-AC52-06NA25396. A.I.N. acknowledges the support from the CONACyT, Grant No. 118930. J.C.B.Z. acknowledges the support from the CONACyT, Grant No. 171014.

\appendix

\section{Exact solution of the Non-Hermitian Landau-Zener problem}

The non-Hermitian Hamiltonian, $ \mathcal {H}_k(t)$, projected on the two-dimensional subspace spanned by $|k_1\rangle= {\scriptsize \left(
                      \begin{array}{c}
                        1 \\
                        0 \\
                      \end{array}
                    \right)}$ and $|k_0\rangle ={\scriptsize \left(
                      \begin{array}{c}
                        0 \\
                        1 \\
                      \end{array}
                    \right)}$, takes the form
\begin{align}\label{AH1}
 \mathcal {H}_k(t)  = -\varepsilon_0(t) {1\hspace{-.125cm}1} +J \left(
\begin{array}{cc}
              \tilde g(t)- \cos \varphi_k & \sin \varphi_k \\
             \sin \varphi_k & -\tilde g(t)+ \cos \varphi_k   \\
            \end{array}
          \right)
\end{align}
where $\varepsilon_0(t)= J\cos \varphi_k+ iJ\delta(t)$ and $\tilde g(t) = g(t) + i\delta(t)$.
We assume a linear dependence of the function, $\tilde g(t)$, on time:
\begin{align}
\tilde g(t) = \left\{
\begin{array}{l}
\gamma (\tau-t), \quad 0 \leq t \leq \tau \\
0 , \quad t > \tau
\end{array}
\right.,
\end{align}
where, $\gamma=(g+i\delta)/\tau$, and $g$, $\delta$ are real parameters.

The general wave functions,  $|\psi_k\rangle$ and  $\langle\tilde \psi_k |$, satisfy the Schr\"odinger equation and its adjoint equation
 \begin{eqnarray}\label{AEqh1}
i\frac{\partial }{\partial t}|\psi_k\rangle&= \mathcal {H}_k(t)|\psi_k\rangle , \\
-i\frac{\partial }{\partial t}\langle\tilde \psi_k |& =
\langle\tilde \psi_k |\mathcal {H}_k(t)\label{NS_2}.
\end{eqnarray}
Presenting $|\psi_k(t)\rangle$ as a linear superposition
\begin{align}\label{AS1}
|\psi_k(t)\rangle = (u_k(t)|k_0\rangle + v_k(t)|k_1\rangle) e^{i\int \varepsilon_0(t)dt},
\end{align}
and inserting (\ref{AS1}) into Eq. (\ref{AEqh1}), we obtain
\begin{align}\label{AS2a}
i\dot  u_k &= J\big(-(\tilde  g - \cos \varphi_k)\,u_k + \sin \varphi_k\, v_k\big), \\
i\dot  v_k &= J\big(\sin \varphi_k\, u_k +(\tilde  g - \cos \varphi_k)\,v_k \big ).
\label{AS2b}
\end{align}

Let $z_k(t) = e^{i\pi/4}\sqrt{2J/\gamma}\big(\gamma(\tau -t)-\cos \varphi_k\big)$ be a new variable. Then, for new functions, $u_k(t)= U_k(z_k)$ and $v_k(t) = V_k(z_k)$, we write Eqs. (\ref{AS2a}), (\ref{AS2b}) in the standard Landau-Zener form \citep{LL,ZC}
\begin{align}\label{AS3a}
\frac{d}{dz_k}U_k  = - \frac{z_k}{2}U_k +\sqrt{i\nu_k}V_k, \\
\frac{d}{dz_k}V_k  =  \frac{z_k}{2}V_k +\sqrt{i\nu_k}U_k,
\label{AS3b}
\end{align}
where $\nu_k= J\sin^2 \varphi_k/2\gamma$, and the complex `time' $z_k$ runs from $z_k(0) = e^{i\pi/4}\sqrt{2J/\gamma}\big(\gamma \tau-\cos \varphi_k\big)$ to $z_k(\tau) = -e^{i\pi/4}\sqrt{2J/\gamma}\cos \varphi_k$.

From Eqs. (\ref{AS3a}), (\ref{AS3b}) we obtain the second order Weber equation
\begin{align}
\frac{d^2}{dz^2_k}U_k  + \Big(\frac{1}{2} - \frac{z_k^2}{4} - i\nu_k\Big)U_k =0 ,\\
\frac{d^2}{dz^2_k}V_k  - \Big(\frac{1}{2} +\frac{z_k^2}{4} +i\nu_k\Big)V_k =0 .
\end{align}
Solution of the Weber's equation is given by the parabolic cylinder functions $D_{-i\nu_k}(\pm z)$, $D_{i\nu_k-1}(\pm i z)$.

We obtain the solutions of Eqs. (\ref{AS3a}, \ref{AS3b}) in the form
 \begin{align}\label{AEq4}
& U_{k}(z_k)= B_k D_{-i\nu_k}(z_k)  -{i}{\sqrt{i\nu_k }}  A_k D_{i\nu_k-1}(iz_k)  ,\\
& V_{k}(z_k)=   A_k D_{i\nu_k}(iz_k) -\sqrt{i\nu_k } B_k D_{-i\nu_k-1}(z_k)  ,
\label{AEq4a}
 \end{align}
where the constants, $A_k$ and $B_k$, are determined from the initial conditions.

If the evolution of TLS starts at $t_0 = 0$ in the `ground' state, $|\psi(0)\rangle = |k_0\rangle$, the following initial conditions should be imposed: $u_k(0)=1$ and $v_k(0)=0$. Using the identity (\ref{ID}), we obtain
\begin{align}
A_k={\sqrt{i\nu_k }}e^{\pi\nu_k/2}  D_{-i\nu_k-1}\big(z_k(0)\big), \\
B_k= e^{\pi\nu_k/2}  D_{i\nu_k}\big(iz_k(0)\big) .
\end{align}

We assume further that $\tau \gg g/J$. This implies $|z_k(0)| \gg 1$. Then, applying the asymptotic formulas for the parabolic cylinder functions with $-3\pi/4 < \arg z_k < 3\pi/4$, we obtain
\begin{align}\label{A3}
 A_k= 0 + {\mathcal O} (1/|z_k(0)|), \\
 B_k=(z_k(0))^{i\nu_k}\,e^{z^2_k(0)/4}.
\end{align}

Similar consideration of the adjoint Schr\"odinger equation
with the wavefunction
\begin{align}\label{AS1b}
\langle\tilde\psi_k | = \big(\tilde u_k(t)\langle k_0| + \tilde v_k(t)\langle k_1|\big) e^{-i\int \varepsilon_0(t)dt},
\end{align}
yields
\begin{align}\label{AS2_b}
i\dot  {\tilde u}_k &= -J\big(-(\tilde  g - \cos \varphi_k)\,{\tilde u}_k + \sin \varphi_k\, {\tilde v}_k\big), \\
i\dot  {\tilde v}_k &= -J\big(\sin \varphi_k\, {\tilde u}_k +(\tilde  g - \cos \varphi_k)\,{\tilde v}_k\big ).
\label{AS_2b}
\end{align}
For the functions, $\tilde u_k(t)=\tilde U_k(z_k)$ and $\tilde v_k(t) = \tilde V_k(z_k)$, we obtain
\begin{align}\label{AS3c}
\frac{d}{dz_k}\tilde U_k  =  \frac{z_k}{2}\tilde U_k - \sqrt{i\nu_k}\tilde V_k, \\
\frac{d}{dz_k}\tilde V_k  = -\frac{z_k}{2}\tilde V_k - \sqrt{i\nu_k}\tilde U_k.
\label{AS3d}
\end{align}
From here it follows
\begin{align}
\frac{d^2}{dz^2_k}\tilde U_k  - \Big(\frac{1}{2} + \frac{z_k^2}{4} + i\nu_k\Big)\tilde U_k =0 ,\\
\frac{d^2}{dz^2_k}\tilde V_k  + \Big(\frac{1}{2} -\frac{z_k^2}{4} -i\nu_k\Big)\tilde V_k =0 .
\end{align}
The solutions are given by
\begin{align}\label{AEq4c}
& \tilde U_{k}(z_k)=  \tilde  A_k D_{-i\nu_k-1}(z_k)+ \tilde  B_k D_{i\nu_k}(iz_k), \\
& \tilde V_{k}(z_k)=  \frac{1}{\sqrt{i\nu_k}} \tilde  A_k D_{-i\nu_k}(z_k) +i
\sqrt{i\nu_k}\tilde B_k D_{i\nu_k-1}(i z_k),
\label{AEq4d}
 \end{align}
where
 \begin{align}
\tilde A_k= \nu_k e^{\pi\nu_k/2}D_{i\nu_k-1}\big(iz_k(0)\big), \\
\tilde B_k= e^{\pi\nu_k/2} D_{-i\nu_k}\big(z_k(0)\big).
\end{align}
 For $|z_k(0)| \gg 1$ we obtain
\begin{align}\label{A13}
\tilde A_k= 0 + {\mathcal O} (1/|z_k(0)|), \\
\tilde B_k=e^{\pi\nu_k/2} (z_k(0))^{-i\nu_k}\,e^{-z^2_k(0)/4}.
\end{align}
Finally, the straightforward computation shows that the obtained solutions satisfy the normalization condition $\langle \tilde \psi_k(t)|\psi_k(t)\rangle = 1$.

The solutions of the Schr\"odinger equations for $|z_k(0)| \gg 1$ and the initial conditions: $u_k(0)=\tilde u_k(0)=1$, $v_k(0)=\tilde v_k(0)=0$, are given by
  \begin{align}\label{AEq5a}
& U_{k}(z_k)= B_k D_{-i\nu_k}(z_k)    ,\\
& V_{k}(z_k)=  -\sqrt{i\nu_k } B_k D_{-i\nu_k-1}(z_k)  ,\\
& \tilde U_{k}(z_k)=   \tilde  B_k D_{i\nu_k}(iz_k), \\
& \tilde V_{k}(z_k)=  i \sqrt{i\nu_k}\tilde B_k D_{i\nu_k-1}(i z_k),
\label{AEq5b}
 \end{align}
where
\begin{align}\label{A_3}
 B_k=(z_k(0))^{i\nu_k}\,e^{z^2_k(0)/4}, \\
 \tilde B_k=e^{\pi\nu_k/2} (z_k(0))^{-i\nu_k}\,e^{-z^2_k(0)/4}.
 \label{A4}
\end{align}

\subsection{Some important properties of the Weber functions}

The parabolic cylinder functions, $D_{-i\nu_k}(\pm z)$, $D_{i\nu_k-1}(\pm i z)$, being solution of the linear differential equation
\begin{align}\label{Cyl}
  \frac{d^2 U }{dz^2} + \Big(\frac{1}{2} + \nu - \frac{z^2}{4} \Big)U =0 ,
\end{align}
satisfy the following derivative and recurrence relations \cite{EMOT1}:
\begin{align}\label{ID1}
 \frac{d }{dz_k}\bigg(e^{z_k^2/4}D_{-i\nu_k}(z_k) \bigg)= i\nu_k e^{z_k^2/4}D_{-i\nu_k-1}(z_k), \\
 \frac{d }{dz_k}\bigg(e^{-z_k^2/4}D_{-i\nu_k}(z_k) \bigg)= - e^{-z_k^2/4}D_{-i\nu_k+1}(z_k).
\end{align}
\begin{align}\label{ID2}
 D_{-i\nu_k+1}(z_k)- z_k D_{-i\nu_k}(z_k) + \nu_k D_{-i\nu_k-1}(z_k)=0.
\end{align}
The Wronskian for these solutions is given by
\begin{align}\label{WR1}
 &D_{-i\nu_k}(z_k)\frac{d }{dz_k} D_{-i\nu_k}(-z_k) \nonumber \\
 &-  D_{-i\nu_k}(-z_k)\frac{d }{dz_k} D_{-i\nu_k}(z_k) = \frac{\sqrt{2\pi}}{\Gamma(i\nu_k)}, \\
 & D_{-i\nu_k}(z_k)\frac{d }{dz_k} D_{i\nu_k+1}(iz_k) \nonumber \\
 &-  D_{i\nu_k+1}(iz_k)\frac{d }{dz_k} D_{-i\nu_k}(z_k) = -ie^{\pi\nu_k/2}.
 \label{WR2}
\end{align}
Using Eqs. (\ref{ID1}) -- (\ref{WR2}), we obtain
\begin{align}\label{ID}
D_{-i\nu_k}(z_k)  D_{i\nu_k}(iz_k)+\nu_k D_{-i\nu_k-1}(z_k)  D_{i\nu_k-1}(iz_k)= e^{-\pi\nu_k/2}
\end{align}

For small value of the argument one can use the power-series expansion of Weber function yielding
\begin{align}\label{W0}
   D_{-i\nu_k}(z_k)= 2^{-i\nu_k/2} \frac{\sqrt{\pi}}{\Gamma\big(\frac{1}{2}+ \frac{i\nu_k}{2}\big)}, \; \; {\rm as}\; |z_k| \rightarrow 0.
\end{align}

\subsubsection{Asymptotic expansion for large value of argument }

For large value of argument, $|z_k| \gg 1$, and for $|\arg z_k |< 3\pi/4$ the following asymptotic expansion is valid \cite{EMOT1}
\begin{align}\label{W1}
  D_{-i\nu_k}(z_k)= z_k^{-i\nu_k}e^{-z_k^2/4}\big(1 + {\mathcal O}(|z_k^2|^{-1} )\big)
\end{align}
To find the asymptotics of the Weber functions for other values of  its argument one can use the relations:
\begin{align}\label{W2}
 D_{-i\nu_k}(z_k)= e^{-\pi\nu_k}D_{-i\nu_k}(-z_k)
 - \frac{i\sqrt{2\pi}}{\Gamma(i\nu_k)}e^{-\pi\nu_k/2}D_{i\nu_k-1}(iz_k).
\end{align}
In particular, for $-5\pi/4 <\arg z_k <-\pi/4$, this yields
\begin{align}\label{W3}
 D_{-i\nu_k}(z_k)=& z_k^{-i\nu_k}e^{-z_k^2/4}\big(1 + {\mathcal O}(|z_k^2|^{-1}) \big) \nonumber\\
 &+ \frac{i\sqrt{2\pi}}{\Gamma(i\nu_k)}e^{-\pi\nu_k}z_k^{i\nu_k-1} e^{z_k^2/4}\big(1 + {\mathcal O}|z_k^2|^{-1} \big).
\end{align}

\subsubsection{Large-order asymptotics }

For large-order value of the Weber functions  with a phase of argument $|\arg z_k |< \pi/2$ the leading terms are \cite{FWO,VGBM}
\begin{align}\label{AW1}
 & D_{-i\nu_k}(z_k)\sim \cos\frac{\theta_k}{2}\,e^{\pi \nu_k/4 - i\eta}\bigg(1 + {\mathcal O} \Big(\frac{1}{\sqrt{|\nu_k|}}\Big)\bigg), \\
&  D_{-i\nu_k-1}(z_k)\sim \frac{1}{\sqrt{i\nu_k}}\sin\frac{\theta_k}{2}\,e^{\pi \nu_k/4 - i\eta}\bigg(1 + {\mathcal O} \Big(\frac{1}{\sqrt{|\nu_k|}}\Big)\bigg),
  \label{AW2}
     \end{align}
where
\begin{align}
\eta =& - \frac{\nu_k}{2} +{\nu_k} \ln\bigg(\frac{1}{2}\Big(z_k e^{-i\pi/4}+ \sqrt{4\nu_k^2 -i z^2_k} \Big) \bigg) \nonumber \\
& + \frac{z_k e^{-i\pi/4}}{4}\sqrt{4\nu_k^2 -i z^2_k},
\end{align}
and
\begin{align}\label{AW3}
\cos\theta_k = \frac{z_k  }{\sqrt{ z^2_k + 4i\nu_k^2 }}.
\end{align}

\section{Equation of motion}

We consider a two-level system governed by the non-Hermitian Hamiltonian, $\tilde{ {\mathcal H}}_{\rm eff}$, written as
\begin{equation}\label{H4a}
\tilde{ {\mathcal H}}_{\rm eff}= \frac{\tilde\lambda_0}{2} {1\hspace{-.17cm}1}+ \frac{1}{2}{ \boldsymbol {\tilde \Omega }}(t)\cdot \boldsymbol \sigma,
\end{equation}
where $\boldsymbol {\tilde\Omega}(t) = (\tilde\Omega_x(t),\tilde\Omega_y(t),\tilde\Omega_z(t))$  is the complex vector and $\tilde\lambda_0 = \lambda_0 -i\Gamma$, where $\Gamma =(\Gamma_0 +\Gamma_1)/2$. The qubit states $|u(t)\rangle$ and $\langle u(t)|$ satisfy the Schr\"odinger equation:
\begin{align}\label{EqB1c}
 i\frac{\partial|u(t)\rangle}{\partial t} =&\tilde{ {\mathcal H}}_{\rm eff} |u(t)\rangle, \\
  -i\frac{\partial\langle u(t)|}{\partial t} =&\langle u(t)|\tilde{ {\mathcal H}}^{\dagger}_{\rm eff}.
  \label{EqB2d}
 \end{align}

Employing Eqs.(\ref{EqB1c}), (\ref{EqB2d}), we find that the Bloch vector, $\mathbf n(t) = \langle u(t)|\boldsymbol\sigma |u(t)\rangle$, satisfies the following generalized Bloch equation (GBE):
\begin{align}\label{EB2}
\frac{d \mathbf n}{dt} = &- \Gamma  \mathbf n  + n\,\Im \boldsymbol { \tilde\Omega }(t) + \Re\boldsymbol { \tilde\Omega }(t)\times \mathbf n,\\
\frac{d n}{dt} = &- \Gamma n + \Im \boldsymbol {\tilde \Omega }(t)\cdot \mathbf n,
\end{align}
where $n = \langle u(t)|u(t)\rangle= \sqrt{n_x^2+n_y^2+n_z^2}$.

Denoting the real part of the complex vector $ {\boldsymbol {\tilde \Omega }}$ as $ {\Re \boldsymbol {\tilde \Omega }} = (\Omega_x,\Omega_y,\Omega_z)$ and its imaginary part as ${\Im \boldsymbol { \Omega }}= (\Lambda_x,\Lambda_y,\Lambda_z)$, we obtain
\begin{align}\label{eqB5b}
   \frac{d n_x}{dt} = &-\Gamma n_x+ \Lambda_x n + \Omega_y n_z - \Omega_z n_y, \\
   \frac{d n_y}{dt} = &-\Gamma n_y +  \Lambda_y n -\Omega_x n_z + \Omega_z n_x,\\
    \frac{d n_z}{dt} = &-\Gamma n_z +  \Lambda_z n  +\Omega_x n_y - \Omega_y n_x, \\
     \frac{d n}{dt} = &- \Gamma n +  \Lambda_x n_x +  \Lambda_y n_y + \Lambda_z n_z .
    \label{eqB5c}
\end{align}

In terms of the Bloch vector the qubit state population of upper/lower level can be written as follows
\begin{align}\label{P6d}
 \rho_{11}(t) = \frac{1}{2}(n(t)  +n_z(t) ) , \\
  \rho_{00}(t) = \frac{1}{2}(n(t)  -n_z(t) ).
  \label{P6_d}
\end{align}
This yields, $n_z(t) = \rho_{11}(t) -  \rho_{00}(t) $ and $n(t) =\rho_{11}(t) +  \rho_{00}(t) $. Substituting these expressions for $n_z(t)$ and $n(t)$ into Eqs. (\ref{eqB5b}) - (\ref{eqB5c}), we obtain
\begin{align}\label{eqB7a}
         \frac{d n_x}{dt} = &-\Gamma n_x +\Lambda_x(\rho_{11} +  \rho_{00})\nonumber \\
         &   + \Omega_y(\rho_{11} -  \rho_{00}) - \Omega_z n_y, \\
   \frac{d n_y}{dt} = &-\Gamma n_y + \Lambda_y(\rho_{11} +  \rho_{00})\nonumber \\
   & -\Omega_x(\rho_{11} - \rho_{00}) + \Omega_z n_x,\\
    \frac{d \rho_{11} }{dt} = &-(\Gamma -\Lambda_z)\rho_{11}+\frac{1}{2}(\Lambda_x n_x + \Lambda_y n_y ) \nonumber \\
    & +\frac{1}{2}(\Omega_x n_y - \Omega_y n_x ),\\
     \frac{d \rho_{00}}{dt} = &-(\Gamma +\Lambda_z) \rho_{00}+ \frac{1}{2}(\Lambda_x n_x + \Lambda_y n_y ) \nonumber\\
     & -\frac{1}{2}(\Omega_x n_y - \Omega_y n_x ),\\
     \frac{d n}{dt} = &- \Gamma n +  \Lambda_x n_x +  \Lambda_y n_y + \Lambda_z (\rho_{11} -  \rho_{00}).
    \label{eqB7b}
\end{align}


\begin{thebibliography}{47}
\expandafter\ifx\csname natexlab\endcsname\relax\def\natexlab#1{#1}\fi
\expandafter\ifx\csname bibnamefont\endcsname\relax
  \def\bibnamefont#1{#1}\fi
\expandafter\ifx\csname bibfnamefont\endcsname\relax
  \def\bibfnamefont#1{#1}\fi
\expandafter\ifx\csname citenamefont\endcsname\relax
  \def\citenamefont#1{#1}\fi
\expandafter\ifx\csname url\endcsname\relax
  \def\url#1{\texttt{#1}}\fi
\expandafter\ifx\csname urlprefix\endcsname\relax\def\urlprefix{URL }\fi
\providecommand{\bibinfo}[2]{#2}
\providecommand{\eprint}[2][]{\url{#2}}

\bibitem[{\citenamefont{Kadowaki and Nishimori}(1998)}]{KN}
\bibinfo{author}{\bibfnamefont{T.}~\bibnamefont{Kadowaki}} \bibnamefont{and}
  \bibinfo{author}{\bibfnamefont{H.}~\bibnamefont{Nishimori}},
  \bibinfo{journal}{Phys. Rev. E} \textbf{\bibinfo{volume}{58}},
  \bibinfo{pages}{5355} (\bibinfo{year}{1998}).

\bibitem[{\citenamefont{Farhi et~al.}(2001)\citenamefont{Farhi, Goldstone,
  Gutmann, Lapan, Lundgren, and Preda}}]{FGGLL}
\bibinfo{author}{\bibfnamefont{E.}~\bibnamefont{Farhi}},
  \bibinfo{author}{\bibfnamefont{J.}~\bibnamefont{Goldstone}},
  \bibinfo{author}{\bibfnamefont{S.}~\bibnamefont{Gutmann}},
  \bibinfo{author}{\bibfnamefont{J.}~\bibnamefont{Lapan}},
  \bibinfo{author}{\bibfnamefont{A.}~\bibnamefont{Lundgren}}, \bibnamefont{and}
  \bibinfo{author}{\bibfnamefont{D.}~\bibnamefont{Preda}},
  \bibinfo{journal}{Science} \textbf{\bibinfo{volume}{292}},
  \bibinfo{pages}{472} (\bibinfo{year}{2001}).

\bibitem[{\citenamefont{Suzuki and Okada}(2005{\natexlab{a}})}]{SSMO}
\bibinfo{author}{\bibfnamefont{S.}~\bibnamefont{Suzuki}} \bibnamefont{and}
  \bibinfo{author}{\bibfnamefont{M.}~\bibnamefont{Okada}}, in
  \emph{\bibinfo{booktitle}{{Quantum Annealing and Related Optimization
  Methods}}}, edited by \bibinfo{editor}{\bibfnamefont{A.}~\bibnamefont{Das}}
  \bibnamefont{and} \bibinfo{editor}{\bibfnamefont{B.~K.}
  \bibnamefont{Chakrabarti}} (\bibinfo{publisher}{Springer},
  \bibinfo{year}{2005}{\natexlab{a}}), vol. \bibinfo{volume}{679} of
  \emph{\bibinfo{series}{Lecture Notes in Physics}}, pp. \bibinfo{pages}{207 --
  238}.

\bibitem[{\citenamefont{Das and Chakrabarti}(2008)}]{DC}
\bibinfo{author}{\bibfnamefont{A.}~\bibnamefont{Das}} \bibnamefont{and}
  \bibinfo{author}{\bibfnamefont{B.~K.} \bibnamefont{Chakrabarti}},
  \bibinfo{journal}{Rev. Mod. Phys.} \textbf{\bibinfo{volume}{80}},
  \bibinfo{eid}{1061} (\bibinfo{year}{2008}).

\bibitem[{\citenamefont{Santoro et~al.}(2002)\citenamefont{Santoro, Martonak,
  Tosatti, and Car}}]{SMTC}
\bibinfo{author}{\bibfnamefont{G.~E.} \bibnamefont{Santoro}},
  \bibinfo{author}{\bibfnamefont{R.}~\bibnamefont{Martonak}},
  \bibinfo{author}{\bibfnamefont{E.}~\bibnamefont{Tosatti}}, \bibnamefont{and}
  \bibinfo{author}{\bibfnamefont{R.}~\bibnamefont{Car}},
  \bibinfo{journal}{Science} \textbf{\bibinfo{volume}{295}},
  \bibinfo{pages}{2427} (\bibinfo{year}{2002}).

\bibitem[{\citenamefont{Ohzeki and Nishimori}(2011{\natexlab{a}})}]{OMNH}
\bibinfo{author}{\bibfnamefont{M.}~\bibnamefont{Ohzeki}} \bibnamefont{and}
  \bibinfo{author}{\bibfnamefont{H.}~\bibnamefont{Nishimori}},
  \bibinfo{journal}{J. Comp. Theor. Nanoscience} \textbf{\bibinfo{volume}{8}},
  \bibinfo{pages}{963} (\bibinfo{year}{2011}{\natexlab{a}}).

\bibitem[{\citenamefont{Tanaka and Tamura}(2012)}]{TSTR}
\bibinfo{author}{\bibfnamefont{S.}~\bibnamefont{Tanaka}} \bibnamefont{and}
  \bibinfo{author}{\bibfnamefont{R.}~\bibnamefont{Tamura}}, in
  \emph{\bibinfo{booktitle}{Lectures on Quantum Computing, Thermodynamics and
  Statistical Physics}}, edited by
  \bibinfo{editor}{\bibfnamefont{M.}~\bibnamefont{Nakahara}} \bibnamefont{and}
  \bibinfo{editor}{\bibfnamefont{S.}~\bibnamefont{Tanaka}}
  (\bibinfo{publisher}{World Scientific}, \bibinfo{year}{2012}),
  vol.~\bibinfo{volume}{8} of \emph{\bibinfo{series}{Kinki University Series on
  Quantum Computing}}, pp. \bibinfo{pages}{3--62}.

\bibitem[{\citenamefont{Stella et~al.}(2005)\citenamefont{Stella, Santoro, and
  Tosatti}}]{SST}
\bibinfo{author}{\bibfnamefont{L.}~\bibnamefont{Stella}},
  \bibinfo{author}{\bibfnamefont{G.~E.} \bibnamefont{Santoro}},
  \bibnamefont{and} \bibinfo{author}{\bibfnamefont{E.}~\bibnamefont{Tosatti}},
  \bibinfo{journal}{Phys. Rev. B} \textbf{\bibinfo{volume}{72}},
  \bibinfo{pages}{014303} (\bibinfo{year}{2005}).

\bibitem[{\citenamefont{Suzuki et~al.}(2007)\citenamefont{Suzuki, Nishimori,
  and Suzuki}}]{SNS}
\bibinfo{author}{\bibfnamefont{S.}~\bibnamefont{Suzuki}},
  \bibinfo{author}{\bibfnamefont{H.}~\bibnamefont{Nishimori}},
  \bibnamefont{and} \bibinfo{author}{\bibfnamefont{M.}~\bibnamefont{Suzuki}},
  \bibinfo{journal}{Phys. Rev. E} \textbf{\bibinfo{volume}{75}},
  \bibinfo{pages}{051112} (\bibinfo{year}{2007}).

\bibitem[{\citenamefont{Amin}(2008)}]{AM}
\bibinfo{author}{\bibfnamefont{M.~H.~S.} \bibnamefont{Amin}},
  \bibinfo{journal}{Phys. Rev. Lett.} \textbf{\bibinfo{volume}{100}},
  \bibinfo{eid}{130503} (\bibinfo{year}{2008}).

\bibitem[{\citenamefont{Smelyanskiy et~al.}()\citenamefont{Smelyanskiy,
  v~Toussaint, and Timucin}}]{SUD}
\bibinfo{author}{\bibfnamefont{V.~N.} \bibnamefont{Smelyanskiy}},
  \bibinfo{author}{\bibfnamefont{U.}~\bibnamefont{v~Toussaint}},
  \bibnamefont{and} \bibinfo{author}{\bibfnamefont{D.~A.}
  \bibnamefont{Timucin}}, \emph{\bibinfo{title}{{Dynamics of quantum adiabatic
  evolution algorithm for Number Partitioning, \rm arXiv: quant-ph/0202155}}}.

\bibitem[{\citenamefont{J\"org et~al.}(2008)\citenamefont{J\"org, Krzakala,
  Kurchan, and Maggs}}]{JKKM}
\bibinfo{author}{\bibfnamefont{T.}~\bibnamefont{J\"org}},
  \bibinfo{author}{\bibfnamefont{F.}~\bibnamefont{Krzakala}},
  \bibinfo{author}{\bibfnamefont{J.}~\bibnamefont{Kurchan}}, \bibnamefont{and}
  \bibinfo{author}{\bibfnamefont{A.~C.} \bibnamefont{Maggs}},
  \bibinfo{journal}{Phys. Rev. Lett.} \textbf{\bibinfo{volume}{101}},
  \bibinfo{eid}{147204} (\bibinfo{year}{2008}).

\bibitem[{\citenamefont{Young et~al.}(2008)\citenamefont{Young, Knysh, and
  Smelyanskiy}}]{YKS}
\bibinfo{author}{\bibfnamefont{A.~P.} \bibnamefont{Young}},
  \bibinfo{author}{\bibfnamefont{S.}~\bibnamefont{Knysh}}, \bibnamefont{and}
  \bibinfo{author}{\bibfnamefont{V.~N.} \bibnamefont{Smelyanskiy}},
  \bibinfo{journal}{Phys. Rev. Lett.} \textbf{\bibinfo{volume}{101}},
  \bibinfo{pages}{170503} (\bibinfo{year}{2008}).

\bibitem[{\citenamefont{Berman and Nesterov}(2009)}]{BN}
\bibinfo{author}{\bibfnamefont{G.~P.} \bibnamefont{Berman}} \bibnamefont{and}
  \bibinfo{author}{\bibfnamefont{A.~I.} \bibnamefont{Nesterov}},
  \bibinfo{journal}{IJQI} \textbf{\bibinfo{volume}{7}}, \bibinfo{pages}{1469}
  (\bibinfo{year}{2009}).

\bibitem[{\citenamefont{Nesterov and Berman}(2012)}]{NABG}
\bibinfo{author}{\bibfnamefont{A.~I.} \bibnamefont{Nesterov}} \bibnamefont{and}
  \bibinfo{author}{\bibfnamefont{G.~P.} \bibnamefont{Berman}},
  \bibinfo{journal}{Phys. Rev. A} \textbf{\bibinfo{volume}{86}},
  \bibinfo{pages}{052316} (\bibinfo{year}{2012}).

\bibitem[{\citenamefont{Berry}(1984)}]{B0}
\bibinfo{author}{\bibfnamefont{M.~V.} \bibnamefont{Berry}},
  \bibinfo{journal}{Proc. R. Soc. A} \textbf{\bibinfo{volume}{392}},
  \bibinfo{pages}{45 } (\bibinfo{year}{1984}).

\bibitem[{\citenamefont{Berry and Wilkinson}(1984)}]{BW}
\bibinfo{author}{\bibfnamefont{M.~V.} \bibnamefont{Berry}} \bibnamefont{and}
  \bibinfo{author}{\bibfnamefont{M.}~\bibnamefont{Wilkinson}},
  \bibinfo{journal}{Proc. Roy. Soc. A} \textbf{\bibinfo{volume}{392}},
  \bibinfo{pages}{15 } (\bibinfo{year}{1984}).

\bibitem[{\citenamefont{Morse and Feshbach}(1953)}]{MF}
\bibinfo{author}{\bibfnamefont{P.~M.} \bibnamefont{Morse}} \bibnamefont{and}
  \bibinfo{author}{\bibfnamefont{H.}~\bibnamefont{Feshbach}},
  \emph{\bibinfo{title}{Methods of Theoretical Physics}}
  (\bibinfo{publisher}{McGraw-Hill}, \bibinfo{address}{New York},
  \bibinfo{year}{1953}).

\bibitem[{\citenamefont{{A. P. Seyranian, O. N. Kirillov and A. A.
  Mailybaev}}(2005)}]{SKM}
\bibinfo{author}{\bibnamefont{{A. P. Seyranian, O. N. Kirillov and A. A.
  Mailybaev}}}, \bibinfo{journal}{J. Phys. A: Math. Gen.}
  \textbf{\bibinfo{volume}{38}}, \bibinfo{pages}{1723 } (\bibinfo{year}{2005}).

\bibitem[{\citenamefont{Kirillov et~al.}(2005)\citenamefont{Kirillov,
  Mailybaev, and Seyranian}}]{KMS}
\bibinfo{author}{\bibfnamefont{O.~N.} \bibnamefont{Kirillov}},
  \bibinfo{author}{\bibfnamefont{A.~A.} \bibnamefont{Mailybaev}},
  \bibnamefont{and} \bibinfo{author}{\bibfnamefont{A.~P.}
  \bibnamefont{Seyranian}}, \bibinfo{journal}{J. Phys. A}
  \textbf{\bibinfo{volume}{38}}, \bibinfo{pages}{5531 } (\bibinfo{year}{2005}).

\bibitem[{\citenamefont{Mailybaev et~al.}(2006)\citenamefont{Mailybaev,
  Kirillov, and Seyranian}}]{MKS1}
\bibinfo{author}{\bibfnamefont{A.~A.} \bibnamefont{Mailybaev}},
  \bibinfo{author}{\bibfnamefont{O.~N.} \bibnamefont{Kirillov}},
  \bibnamefont{and} \bibinfo{author}{\bibfnamefont{A.~P.}
  \bibnamefont{Seyranian}}, \bibinfo{journal}{Doklady Math.}
  \textbf{\bibinfo{volume}{73}}, \bibinfo{pages}{129} (\bibinfo{year}{2006}).

\bibitem[{\citenamefont{Katsura}(1962)}]{KSH}
\bibinfo{author}{\bibfnamefont{S.}~\bibnamefont{Katsura}},
  \bibinfo{journal}{Phys. Rev.} \textbf{\bibinfo{volume}{127}},
  \bibinfo{pages}{1508} (\bibinfo{year}{1962}).

\bibitem[{\citenamefont{Dziarmaga}(2005)}]{DJ}
\bibinfo{author}{\bibfnamefont{J.}~\bibnamefont{Dziarmaga}},
  \bibinfo{journal}{Phys. Rev. Lett.} \textbf{\bibinfo{volume}{95}},
  \bibinfo{eid}{245701} (\bibinfo{year}{2005}).

\bibitem[{\citenamefont{Carollo and Pachos}(2005)}]{CP}
\bibinfo{author}{\bibfnamefont{A.~C.~M.} \bibnamefont{Carollo}}
  \bibnamefont{and} \bibinfo{author}{\bibfnamefont{J.~K.}
  \bibnamefont{Pachos}}, \bibinfo{journal}{Phys. Rev. Lett.}
  \textbf{\bibinfo{volume}{95}}, \bibinfo{pages}{157203}
  (\bibinfo{year}{2005}).

\bibitem[{\citenamefont{Nesterov and Ovchinnikov}(2008)}]{NO}
\bibinfo{author}{\bibfnamefont{A.~I.} \bibnamefont{Nesterov}} \bibnamefont{and}
  \bibinfo{author}{\bibfnamefont{S.~G.} \bibnamefont{Ovchinnikov}},
  \bibinfo{journal}{Phys. Rev. E} \textbf{\bibinfo{volume}{78}},
  \bibinfo{pages}{015202(R)} (\bibinfo{year}{2008}).

\bibitem[{\citenamefont{Lieb et~al.}(1961)\citenamefont{Lieb, Schultz, and
  Mattis}}]{LSM}
\bibinfo{author}{\bibfnamefont{E.}~\bibnamefont{Lieb}},
  \bibinfo{author}{\bibfnamefont{T.}~\bibnamefont{Schultz}}, \bibnamefont{and}
  \bibinfo{author}{\bibfnamefont{D.}~\bibnamefont{Mattis}},
  \bibinfo{journal}{Ann. Phys.} \textbf{\bibinfo{volume}{16}},
  \bibinfo{pages}{407} (\bibinfo{year}{1961}).

\bibitem[{\citenamefont{Lebedev}(1972)}]{Leb}
\bibinfo{author}{\bibfnamefont{M.~N.} \bibnamefont{Lebedev}},
  \emph{\bibinfo{title}{{Special Functions {$\&$} Their Applications}}}
  (\bibinfo{publisher}{Dover}, \bibinfo{address}{New York},
  \bibinfo{year}{1972}).

\bibitem[{\citenamefont{Landau and Lifshitz}(1958)}]{LL}
\bibinfo{author}{\bibfnamefont{L.}~\bibnamefont{Landau}} \bibnamefont{and}
  \bibinfo{author}{\bibfnamefont{E.~M.} \bibnamefont{Lifshitz}},
  \emph{\bibinfo{title}{"Quantum Mechanics"}} (\bibinfo{publisher}{Pergamon},
  \bibinfo{address}{New York}, \bibinfo{year}{1958}).

\bibitem[{\citenamefont{Zener}(1932)}]{ZC}
\bibinfo{author}{\bibfnamefont{C.}~\bibnamefont{Zener}},
  \bibinfo{journal}{Proc. R. Soc. A} \textbf{\bibinfo{volume}{137}},
  \bibinfo{pages}{696} (\bibinfo{year}{1932}).

\bibitem[{\citenamefont{Berry}(1990)}]{BMV1}
\bibinfo{author}{\bibfnamefont{M.~V.} \bibnamefont{Berry}},
  \bibinfo{journal}{Proc. Roy. Soc. A} \textbf{\bibinfo{volume}{430}},
  \bibinfo{pages}{405} (\bibinfo{year}{1990}).

\bibitem[{\citenamefont{Joye et~al.}(1991)\citenamefont{Joye, Mileti, and
  Pfister}}]{JMPC}
\bibinfo{author}{\bibfnamefont{A.}~\bibnamefont{Joye}},
  \bibinfo{author}{\bibfnamefont{G.}~\bibnamefont{Mileti}}, \bibnamefont{and}
  \bibinfo{author}{\bibfnamefont{C.-E.} \bibnamefont{Pfister}},
  \bibinfo{journal}{Phys. Rev. A} \textbf{\bibinfo{volume}{44}},
  \bibinfo{pages}{4280} (\bibinfo{year}{1991}).

\bibitem[{\citenamefont{A.~Joye and Pfister}(1991)}]{JKP}
\bibinfo{author}{\bibfnamefont{H.~K.} \bibnamefont{A.~Joye}} \bibnamefont{and}
  \bibinfo{author}{\bibfnamefont{C.~E.} \bibnamefont{Pfister}},
  \bibinfo{journal}{Ann. Phys.} \textbf{\bibinfo{volume}{208}},
  \bibinfo{pages}{299} (\bibinfo{year}{1991}).

\bibitem[{\citenamefont{Kvitsinsky and Putterman}(1991)}]{KP}
\bibinfo{author}{\bibfnamefont{A.}~\bibnamefont{Kvitsinsky}} \bibnamefont{and}
  \bibinfo{author}{\bibfnamefont{S.}~\bibnamefont{Putterman}},
  \bibinfo{journal}{J. Math. Phys.} \textbf{\bibinfo{volume}{32}},
  \bibinfo{pages}{1403} (\bibinfo{year}{1991}).

\bibitem[{\citenamefont{Schilling et~al.}(2006)\citenamefont{Schilling,
  Vogelsberger, and Garanin}}]{SVG}
\bibinfo{author}{\bibfnamefont{R.}~\bibnamefont{Schilling}},
  \bibinfo{author}{\bibfnamefont{M.}~\bibnamefont{Vogelsberger}},
  \bibnamefont{and} \bibinfo{author}{\bibfnamefont{D.~A.}
  \bibnamefont{Garanin}}, \bibinfo{journal}{J. Phys. A}
  \textbf{\bibinfo{volume}{39}}, \bibinfo{pages}{13727} (\bibinfo{year}{2006}).

\bibitem[{\citenamefont{Dridi et~al.}()\citenamefont{Dridi, Gu\'erin, Jauslin,
  Viennot, and Jolicard}}]{DG}
\bibinfo{author}{\bibfnamefont{G.}~\bibnamefont{Dridi}},
  \bibinfo{author}{\bibfnamefont{S.}~\bibnamefont{Gu\'erin}},
  \bibinfo{author}{\bibfnamefont{H.~R.} \bibnamefont{Jauslin}},
  \bibinfo{author}{\bibfnamefont{D.}~\bibnamefont{Viennot}}, \bibnamefont{and}
  \bibinfo{author}{\bibfnamefont{G.}~\bibnamefont{Jolicard}},
  \bibinfo{journal}{Phys. Rev. A} \textbf{\bibinfo{volume}{82}},
  \bibinfo{pages}{022109} (\bibinfo{year}{2010}).


\bibitem[{\citenamefont{Dykhne}(1962)}]{DAM}
\bibinfo{author}{\bibfnamefont{A.~M.} \bibnamefont{Dykhne}},
  \bibinfo{journal}{Sov. Phys.-JETP} \textbf{\bibinfo{volume}{14}},
  \bibinfo{pages}{941} (\bibinfo{year}{1962}).

\bibitem[{\citenamefont{Davis and Pechukas}(1976)}]{DJP}
\bibinfo{author}{\bibfnamefont{J.~P.} \bibnamefont{Davis}} \bibnamefont{and}
  \bibinfo{author}{\bibfnamefont{P.}~\bibnamefont{Pechukas}},
  \bibinfo{journal}{J. Chem. Phys.} \textbf{\bibinfo{volume}{64}},
  \bibinfo{pages}{3129} (\bibinfo{year}{1976}).

\bibitem[{\citenamefont{Hwang and Pechukas}(1977)}]{HJP}
\bibinfo{author}{\bibfnamefont{J.-T.} \bibnamefont{Hwang}} \bibnamefont{and}
  \bibinfo{author}{\bibfnamefont{P.}~\bibnamefont{Pechukas}},
  \bibinfo{journal}{J. Chem. Phys.} \textbf{\bibinfo{volume}{67}},
  \bibinfo{pages}{4640} (\bibinfo{year}{1977}).

\bibitem[{\citenamefont{Suzuki and Okada}(2005{\natexlab{b}})}]{SSMOK}
\bibinfo{author}{\bibfnamefont{S.}~\bibnamefont{Suzuki}} \bibnamefont{and}
  \bibinfo{author}{\bibfnamefont{M.}~\bibnamefont{Okada}}, \bibinfo{journal}{J.
  Phys. Soc. Japan} \textbf{\bibinfo{volume}{74}}, \bibinfo{pages}{1649}
  (\bibinfo{year}{2005}{\natexlab{b}}).

\bibitem[{\citenamefont{Caneva et~al.}(2007)\citenamefont{Caneva, Fazio, and
  Santoro}}]{CFS}
\bibinfo{author}{\bibfnamefont{T.}~\bibnamefont{Caneva}},
  \bibinfo{author}{\bibfnamefont{R.}~\bibnamefont{Fazio}}, \bibnamefont{and}
  \bibinfo{author}{\bibfnamefont{G.~E.} \bibnamefont{Santoro}},
  \bibinfo{journal}{Phys. Rev. B} \textbf{\bibinfo{volume}{76}},
  \bibinfo{pages}{144427} (\bibinfo{year}{2007}).

\bibitem[{\citenamefont{Erd\'{e}lyi et~al.}(1953)\citenamefont{Erd\'{e}lyi,
  Magnusand, and Oberhettinger}}]{EMOT1}
\bibinfo{author}{\bibfnamefont{A.}~\bibnamefont{Erd\'{e}lyi}},
  \bibinfo{author}{\bibfnamefont{W.}~\bibnamefont{Magnusand}},
  \bibnamefont{and}
  \bibinfo{author}{\bibfnamefont{F.}~\bibnamefont{Oberhettinger}},
  \emph{\bibinfo{title}{{Higher Transcendental Functions}}},
  vol.~\bibinfo{volume}{I} (\bibinfo{publisher}{McGraw-Hill},
  \bibinfo{address}{New York}, \bibinfo{year}{1953}).

\bibitem[{\citenamefont{Zurek et~al.}(2005)\citenamefont{Zurek, Dorner, and
  Zoller}}]{ZDZ}
\bibinfo{author}{\bibfnamefont{W.~H.} \bibnamefont{Zurek}},
  \bibinfo{author}{\bibfnamefont{U.}~\bibnamefont{Dorner}}, \bibnamefont{and}
  \bibinfo{author}{\bibfnamefont{P.}~\bibnamefont{Zoller}},
  \bibinfo{journal}{Phys. Rev. Lett.} \textbf{\bibinfo{volume}{95}},
  \bibinfo{eid}{105701} (pages~\bibinfo{numpages}{4}) (\bibinfo{year}{2005}).

\bibitem[{\citenamefont{Roland and Cerf}(2002)}]{RN}
\bibinfo{author}{\bibfnamefont{J.}~\bibnamefont{Roland}} \bibnamefont{and}
  \bibinfo{author}{\bibfnamefont{N.~J.} \bibnamefont{Cerf}},
  \bibinfo{journal}{Phys. Rev. A} \textbf{\bibinfo{volume}{65}},
  \bibinfo{pages}{042308} (\bibinfo{year}{2002}).

\bibitem[{\citenamefont{Ohzeki}(2010)}]{MO}
\bibinfo{author}{\bibfnamefont{M.}~\bibnamefont{Ohzeki}},
  \bibinfo{journal}{Phys. Rev. Lett.} \textbf{\bibinfo{volume}{105}},
  \bibinfo{pages}{050401} (\bibinfo{year}{2010}).

\bibitem[{\citenamefont{Ohzeki and Nishimori}(2011{\natexlab{b}})}]{ON}
\bibinfo{author}{\bibfnamefont{M.}~\bibnamefont{Ohzeki}} \bibnamefont{and}
  \bibinfo{author}{\bibfnamefont{H.}~\bibnamefont{Nishimori}},
  \bibinfo{journal}{Comp. Phys. Com.} \textbf{\bibinfo{volume}{182}},
  \bibinfo{pages}{257} (\bibinfo{year}{2011}{\natexlab{b}}).

\bibitem[{\citenamefont{Olver}(1959)}]{FWO}
\bibinfo{author}{\bibfnamefont{F.~W.~J.} \bibnamefont{Olver}},
  \bibinfo{journal}{J. Res. Natl. Bur. Stand. B} \textbf{\bibinfo{volume}{63}},
  \bibinfo{pages}{131 } (\bibinfo{year}{1959}).

\bibitem[{\citenamefont{Vitanov and Garraway}(1996)}]{VGBM}
\bibinfo{author}{\bibfnamefont{N.~V.} \bibnamefont{Vitanov}} \bibnamefont{and}
  \bibinfo{author}{\bibfnamefont{B.~M.} \bibnamefont{Garraway}},
  \bibinfo{journal}{Phys. Rev. A} \textbf{\bibinfo{volume}{53}},
  \bibinfo{pages}{4288} (\bibinfo{year}{1996}).

  \bibitem{SU}
  S. Utsunomiya, K. Takata, and Y. Yamamoto, Optics Express, {\bf 19}, 18091 (2011).

  \bibitem{KT}
  K. Takata, S. Utsunomiya, and Y. Yamamoto, New J. Phys. {\bf 14}, 013052 (2012).





\end{thebibliography}
\end{document}